\documentclass[12pt]{JHEP3} 

\usepackage{amsmath,amssymb}
\usepackage{yfonts}
\usepackage{graphicx}
\usepackage{subfigure}
\newcommand\fverb{\setbox\fverbbox=\hbox\bgroup\verb}
\newcommand\fverbdo{\egroup\medskip\noindent%
			\fbox{\unhbox\fverbbox}\ }
\newcommand\fverbit{\egroup\item[\fbox{\unhbox\fverbbox}]}
\newbox\fverbbox

\def\be{\begin{equation}}
\def\ee{\end{equation}}
\def\barr{\begin{array}}
\def\earr{\end{array}}
\def\bea{\begin{eqnarray}}
\def\eea{\end{eqnarray}}
\def\ba{\begin{eqnarray}}
\def\ea{\end{eqnarray}}
\def\a{\alpha}

\title{Holographic Hadrons in a Confining Finite Density Medium}

\author{Yunseok Seo\footnote{yseo@sogang.ac.kr}, Jonathan P. Shock\footnote{shock@fpaxp1.usc.es}, Sang-Jin Sin\footnote{sjsin@hanyang.ac.kr} and Dimitrios Zoakos \footnote{zoakos@fpaxp1.usc.es}
\\
 
$^\ast$Center for Quantum Spacetime, Sogang University, Seoul 121-742, Korea\\

$^\dag$ $^\S$ Departamento de F\'\i sica de Part\'\i culas, Universidade de Santiago de Compostela 
and Instituto Galego de F\'\i sica de Altas Enerx\'\i as (IGFAE)\\
E-15782 Santiago de Compostela, Spain\\

$^\ddag$ Department of Physics, Hanyang University, Seoul 133-791, Korea
}

\abstract{We study a sector of the hadron spectrum in the presence of finite baryon density.
We use a non-supersymmetric gravity dual to a confining guage theory which exhibits a running dilaton.   
The interaction of mesons with the finite density medium is encoded in the dual theory by a
force balancing between flavor D7-branes and a baryon vertex provided by a wrapped D5-brane.
When the current quark mass $m_q$ is sufficiently large, the meson mass reduces, exhibiting an interesting spectral flow as we increase the 
baryon density while it has a more complicated behaviour for very small $m_q$.}

\keywords{AdS/CFT, mesons, baryons, finite density, confinement}

\begin{document} 


\section{Introduction}

Quantum chromodynamics (QCD), known to be the microscopic explanation for the nuclear force,
has been studied for more than 30 years. Nevertheless, reliable and fast methods for treating its dynamics in the low energy regime are still lacking. Although  lattice QCD is rapidly developing, there are problems which this technique is not well suited to, in particular real-time calculations and calculations at finite density due to the infamous sign problem.
We therefore need new ideas which work at least qualitatively in the presence of a chemical potential.  
Recently, techniques derived from string theory have made a remarkable connection between strongly 
interacting gauge theories and gravity in asymptotically Anti De Sitter geometies \cite{Maldacena}. 
Therefore it is natural to ask whether AdS/CFT can shed light on QCD dynamics,  especially in the case of a dense medium. 
Such questions in the absence of finite baryon density have been partially answered in the approach 
known as holographic QCD \cite{SS,EKSS,PR}. Such avenues continue to provide new insights and new possibilities in the realm of understanding real QCD.
The holographic encoding of the baryon chemical potential was studied in 
\cite{FIRST,density,Kobayashi:2006sb}. The equations of state were  analyzed 
in \cite{density} and the spectral functions were calculated in \cite{Holospectral}.
While the prescription for encoding finite temperature has been clear from the early days of AdS/CFT, it is still not clear in all contexts about the geometry describing finite baryon density. This is especially true in the case of QCD-like theories as will become apparent in this work.

The introduction of fundamental matter by means of probe branes was pioneered in 
\cite{Karch:2002sh}. Such an addition allows for the study of meson phenomenology and the response of the system to the baryon density, encoded in a non-zero gauge field configuration on the flavor brane, is seen both in the flavor brane embedding along with the change in the spectral function.

In the deconfined phase it has been shown that when quark density is introduced into the gauge theory (as encoded in a finite density of fundamental strings on the flavor brane worldvolume) the probe brane must end on the black hole horizon \cite{Kobayashi:2006sb}. However, as pointed out in \cite{Seo:2008qc} when a compact D-brane corresponding to a baryon vertex is present (which is not possible in the deconfined geometry), a more natural situation is for the flavor brane to attach to the wrapped baryon 
vertex brane. In the same paper,  the density dependence of the baryon mass was examined in detail. 

In this paper, we consider the meson spectrum  in the presence of finite baryon density in a confining 
gauge theory  following the approach developed in \cite{Seo:2008qc}.
We use the dual confining background at zero temperature found in \cite{Gubser:1999pk} with the addition of flavor D7-branes and a D5-brane baryon vertex joined by a force balancing condition.
We examine the simplest mode on the D7-brane which describes the goldstone boson in the absence of a current quark mass $m_q$.  We find that if the current quark mass $m_q$ is sufficiently large, 
the meson mass reduces as we increase the baryon density while it exhibits a more complicated 
behaviour for small $m_q$, which is similar to the behaviour of a baryon mass in the medium. 
In all cases we find that the baryon density causes a splitting in the spectrum and an intricate spectral flow which can be understood by studying the Schrodinger potential.

The rest of the paper consists of the following sections. 
In section 2 the problem is set up by reviewing baryons and baryon density in holographic QCD. 
In section 3 the interaction between the meson and the baryonic medium is discussed. 
In section 4 the flow of the meson spectrum as a function of the baryon density is shown. 
In section 5 we provide a discussion of the results provided here and add possible lines for future work.

 
\section{Baryons and baryon density in holographic QCD}


\subsection{The confining geometry in  holographic QCD}

We provide here a quick review of the Gubser dilaton flow geometry, \cite{Gubser:1999pk}, which we use as the holographic dual of our confining background. 
Similar non-supersymmetric geometries can be found in 
\cite{Kehagias:1999tr} and \cite{Constable:1999ch}.

The Gubser dilaton flow geometry (GDFG) is a non-supersymmetric deformation of $AdS_5\times S^5$ which corresponds on the field theory side to adding a vev for the supersymmetry breaking term $\left<\rm{tr} \,F^2\right>$ \footnote{It should be noted that pure ${\cal N}=4$ SYM contains no such $S0(6)$ preserving operators but a deformation where an $S0(6)$ invariant mass term for the scalars has been added may allow for such a term}. This operator is sourced on the supergravity side by the dilaton and because of the broken conformal invariance, the running of this operator is seen in the non-trivial profile for the dilaton. In the UV limit we recover the pure $AdS$ geometry but in the IR the theory runs to a strongly coupled (in $g_s$) limit at which point the dilaton blows up, leaving a naked singularity.  By probing this geometry with a fundamental string stretched from the boundary, corresponding to a Wilson loop it can be shown that the geometry presents an area law for the energy of the Wilson loop and therefore exhibits confinement. A mass gap is also found by calculating perturbations of the dilaton\footnote{A further problem is that there is no separation of scales between $R$-charged and non-charged glueball excitations}. The position of the curvature singularity provides an IR scale in the field theory which we can think of as $\Lambda_{QCD}$ which sets all further scales in the problem. The solution has a single dimensionful parameter which dictates the radius of the singularity. We can set this unique dimensionful parameter on the field theory side to 1 thereby scaling all dimensionful parameters in units of $\Lambda_{QCD}$. One clear problem in constructing a realistic theory from such a geometry is that there is no hierarchy between the strong coupling scale and the supersymmetry breaking scale. However, for the purposes of this investigation we will simply utilise the fact that this theory has confinement and a mass gap, just as in QCD.

We clearly cannot trust the geometry in the region around the singularity but for the 
probes we will use to study hadronic physics the singular region exhibits a repulsive potential and we can exclude all solutions where branes end on the singularity. Moreover, according to the criteria of 
\cite{Maldacena:2000mw} and \cite{Gubser:2000nd}, 
this singularity is good. The analysis in \cite{Maldacena:2000mw} concludes that the criterion is that the time component of the metric should not blow up at the singularity. Using the Einstein frame for this analysis, it is easy to prove that the GDFG geometry fulfills this criterion. 

The GDFG is a solution of the type IIB supergravity equations of motion and the solution can be written as:
\be\label{ggmetric}
ds_{10}^2 =e^{\Phi/2}\left(\frac{r^2}{R^2} A^2(r)\eta_{\mu\nu} dx^{\mu}dx^{\nu}
+\frac{R^2}{r^2} dr^2 +R^2 d\Omega_5^2 \right),
\ee
where $A(r)$ and the dilaton are given by:
\be
A(r)=\left(1-\left(\frac{r_{0}}{r}\right)^8 \right)^{1/4}  \qquad \& \qquad 
e^{\Phi}=\left(\frac{(r/r_0)^4+1}{(r/r_0)^4-1}\right)^{\sqrt{3/2}} \, ,
\ee
while the five-form remains unaltered from the pure $AdS$ solution. 
$R$ is the $AdS$ radius and $r_0$ is the position of the singularity which we set to 1 in the following.

The field theory dual of this geometry clearly has some strong similarities to QCD although there are some important differences. We would like simply to exploit the fact that we have a supergravity dual to a confining gauge theory, which is somewhat more realistic than the hard-wall models (see \cite{EKSS} and references therein) in order to further explore QCD-like behaviour. Some of the questions we might ask of such a model are:
\begin{itemize}
\item Can we add flavor to such a model (both quenched and eventually unquenched)?
\item What is the finite temperature description of such a gauge theory?
\item What is the gravity dual of finite baryon density in a confining gauge theory and what is its phenomenology?
\end{itemize}

Section \ref{sec.flavor} reviews the addition of quenched fundamental matter in such a gauge theory  and it will be clear that some qualitatively QCD-like phenomenology comes out of such a picture. The unquenched calculation has not been studied but would certainly be an interesting direction for future work (see \cite{Bigazzi:2009bk} for recent work on studying unquenched flavor in QCD-like models).

In \cite{Kim:2007qk}  and separately in \cite{Evans:2008tu} the finite temperature counterparts to this system were studied, concluding that at finite temperature the only possible solution to the supergravity equations with a horizon had to have a trivial dilation, and therefore no vev for $\left<\rm{tr}\,F^2\right>$. This calculation however does not allow for the possibility of a solution with a naked singularity in the supergravity limit.

The last question above is precisely what we would like to tackle in the current work. In a non-confining geometry one can add finite baryon density to a system simply with the addition of more quarks than antiquarks. We shall deal with this in section \ref{sec.bardens}. However, the problem is more complicated in a confining geometry where we have to introduce a new object - the baryon vertex - in order to accomplish this. Clearly the addition of free quarks does not make sense in such a context.


\subsection{Flavor in the confining geometry}\label{sec.flavor}

For discussions of flavor in the quenched approximation in AdS/CFT we refer the reader to \cite{Karch:2002sh,Kruczenski:2003be,Tedder:2008ug,Erdmenger:2007cm}.

The addition of flavor has been studied extensively in the dilaton flow geometry in \cite{Babington:2003vm,Evans:2004ia,Ghoroku:2004sp}. Here we remind the reader of the most important phenomenological features of a such a setup. We will be able to contrast these with the behaviour of the theory in the presence of a baryon vertex in the forthcoming sections.

For convenience we write the metric (\ref{ggmetric}) in a form where the natural embedding of a D7-brane is clearest. The radial direction and five-sphere combine to form an $\mathbb{E}^{6}$ (up to conformal scalings), written as $\mathbb{E}^4\times\mathbb{E}^2$. The D7-brane lives perpendicular to the $\mathbb{E}^2$ leading to a global $U(1)_R$ symmetry on the worldvolume in the case of massless flavors. The breaking of this $U(1)$ by non-trivial embeddings of the D7-brane corresponds on the field theory side to the breaking of the chiral symmetry through the dynamical generation of a $\left<\bar{q}q\right>$ vacuum expectation value. 

We rewrite the metric as:
\be
ds_{10}^2 =e^{\Phi/2}\left[\frac{r^2}{R^2} A^2(r)\eta_{\mu\nu} dx^{\mu}dx^{\nu}
+\frac{R^2}{r^2}\left(d\rho^2 + \rho^2 d\Omega_3^2 +dL^2 +L^2 d\varphi^2 \right) \right],
\ee
where the $\mathbb{E}^4$ and $\mathbb{E}^2$ are manifest, $r^2 =\rho^2 + L^2$ and $L$ and $\varphi$ are the transverse directions to the D7-brane. Because of the manifest $SO(2)$ symmetry of the geometry we are free to chose a solution with $\varphi=0$. The induced metric on a D7-brane with Euclidean signature is:
\be
ds_{D7}^2 =e^{\Phi/2}\left[\frac{r^2}{R^2} A^2(r)(dt^2+ d\vec{x}^{2})
+\frac{R^2}{r^2}\left\{(1+\dot{L}^2 )d\rho^2 + \rho^2 d\Omega_3^2 \right\} \right],
\ee
where $\dot{L}=dL/d\rho$. In order to study finite baryon density we will be interested in turning on a  non-
vanishing gauge field of the form $A_0(\rho)$.\par
The DBI action for the $N_f$ D7-branes is:
\bea
S_{D7}&=& -N_f \mu_7r_0^4 \int d\xi^8 e^{-\Phi}\sqrt{{\rm det}(g+2\pi \a' F)}\cr
&=& -N_f\tau_7 \int dt d\rho A(r)^3 \rho^3 e^{\Phi/2}\sqrt{e^{\Phi} A(r)^2 (1+\dot{L}^2) -\tilde{F}^2}\, ,
\eea
where we have denoted:
\be
\tau_7 = \mu_7 \Omega_3 V_3,~~~~~\tilde{F} = 2 \pi \a' F_{\rho t}\, ,
\ee
and we have turned the radial variables $\rho$ and $L$ into their dimensionless counterparts via $(\rho,L)
\rightarrow (r_0\rho, r_0 L)$. We do not relabel these variables with a tilde as is often the case, for 
notational simplicity but in the future all dimensionful quantities originating from the holographic energy/
radius duality will be in units of $r_0$ when not explicitely stated. For the current illustration we will set the 
gauge field to zero. How to consistently turn this field on in a confining geometry will be the task of the bulk 
of this paper. At the classical embedding level we have a single equation of motion for the field $L$. In the 
large $\rho$, UV limit the solution has two free parameters, one normalisable and one non-normalisable:
\begin{equation}
L(\rho)_{\rho\rightarrow\infty}=m+\frac{c}{\rho^2}\, .
\end{equation}
These correspond to the source and vacuum expectation value for $\left<\bar{q}q\right>$ (the quark mass $m_q=r_0m/2\pi\alpha'$ and bilinear condensate $\left<\bar{q}q\right>\sim N_c\lambda r_0^3 c$\footnote{for a description of the supersymmetrically complete operator corresponding to $c$ see \cite{Kobayashi:2006sb}}). Although there are two free parameters, the dynamics of the theory in the IR only allows well behaved solutions for a finite number of values of $c$ for a given $m$, in the present case there is just one physical value of $c$ for any $m$. In practice we work backwards and integrate out from the IR with the well behaved solutions given by $L(0)=L_{c}, \,\, L'(0)=0$. 
At the boundary we read off the values of $m$ and $c$. These are parameterised by the one parameter 
family of solutions given by the value $L_{c}$.

In the following figures we show the embeddings for a range of $L_{c}$ values and the plot of the condensate versus the mass calculated parametrically in $L_{c}$.
\begin{figure}[!ht]
\begin{center}
{\includegraphics[angle=0, width=0.8\textwidth]{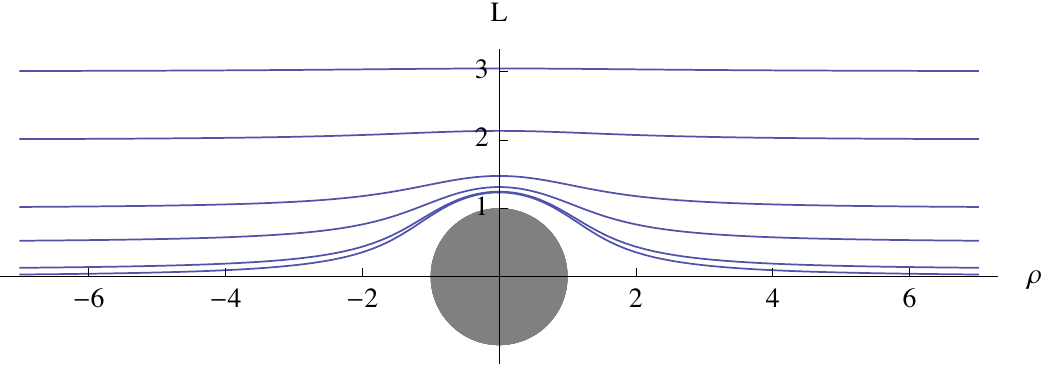}}
\caption{D7-brane embeddings with different values of the IR boundary condition $L(\rho=0)=L_{c}$. In this and all future such plots, the grey disk labels the singular region of the geometry.\label{fig:D7s-01}
}
\end{center}
\end{figure}
\begin{figure}[!ht]
\begin{center}
{\includegraphics[angle=0, width=0.7\textwidth]{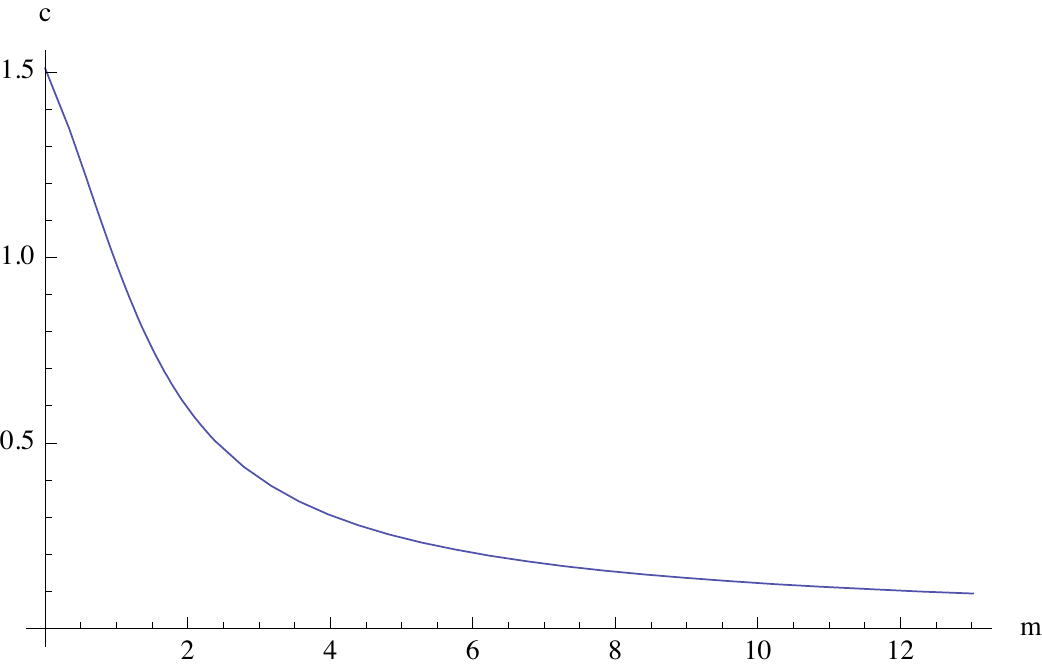}}
\caption{$m$ versus $c$ indicating spontaneous chiral symmetry breaking in the massless limit.\label{fig:mvsc0}
}
\end{center}
\end{figure}
We see that there is spontaneous chiral symmetry breaking and therefore one may expect a goldstone mode. This can be found by studying the excitations of the field in the direction perpendicular to the condensate (as expected from goldstone's theorem). In the case of the D7-brane there are two scalar modes and an 8-compontent gauge field (before gauge fixing) which can be excited on the worldvolume. On top of the classical embedding we can study excitations which correspond to a higgs-like mode, $L$=$L_{emb}+2\pi\alpha'\tilde{L}$. The goldstone mode is given by excitations around the classical solution which in this case has been set to zero by the underlying symmetry: $\varphi=0+2\pi\alpha'\tilde{\phi}$.  We will ignore fields with indices in the three-sphere directions, which are dual to R-charge currents, and set the gauge field compontent $A_\rho$ to zero as a gauge fixing condition. This leaves us with four vector fields. In the case that we study excitations with zero momentum in the Minkowski space, the $S^3$ rotational symmetry leaves us with only two distinct gauge field components, $A_0$ and $A_i$, $i=1..3$. For the moment the interest lies in the Goldstone mode and we remind the reader of the result from \cite{Evans:2004ia} for the behaviour of the goldstone boson mass as a function of the quark mass. This follows the Gell-Mann-Oakes-Renner relation which can be shown exactly by using an expansion about the massless embedding (see \cite{Kruczenski:2003uq,Filev:2009xp} for details).


\subsection{Quark density and chiral symmetry}\label{sec.bardens}

Now we turn to the study of finite quark/baryon density. This has been studied extensively both in the case 
of supersymmetric ${\cal N}=4$ SYM \cite{Karch:2007br} plus flavor and the finite temperature 
counterpart \cite{FIRST,density,Kobayashi:2006sb}.

Finite baryon density corresponds to a non-zero expectation value for the operator $\bar{q}\gamma^0q$ 
which is sourced by the time component of the gauge field on the D7-brane. The solution to such a field 
$A_0$ will generically have a UV behaviour containing both a normalisable and a non-normalisable 
piece. The non-normalisable piece is the source for the above operator which corresponds to the chemical 
potential, whilst the coefficient of the normalisable term is related to the baryon density itself. 

For illustrative purposes we will first discuss the addition of finite baryon density in the case where there is 
a horizon in our space (including both the zero temperature, extremal case and the finite temperature non-
extremal case). At this point it is perhaps more sensible to talk of a finite quark density which is related to 
the baryon density simply by a factor of the number of colors. 
On the D7-brane the non-trivial gauge field configuration acts to deform the brane, especially in the IR and 
the deformation manifests itself as a throat on the brane which ends on the horizon. By studying the 
tension of this throat it can be shown that it corresponds to a bundle of fundamental strings, pulling the D7-
brane into the horizon, where the charges can be swallowed. Such a fundamental density of strings is 
expected due to the finite quark density, the charges of which have to end somewhere and not on the 
vanishing three sphere which the D7-brane otherwise wraps. Thus the brane behaviour is completely 
consistent with the field theory picture. 

The question of interest in the current paper is what happens to a confining gauge theory when finite quark density is added. In the dilaton flow geometry discussed in the previous sections, a D7-brane embedding which falls into the naked singularity would clearly not be a good solution. We can then ask if there is any other configuration which is well defined in the presence of finite quark density.

\begin{figure}[!ht]
\begin{center}
{\includegraphics[angle=0, width=0.7\textwidth]{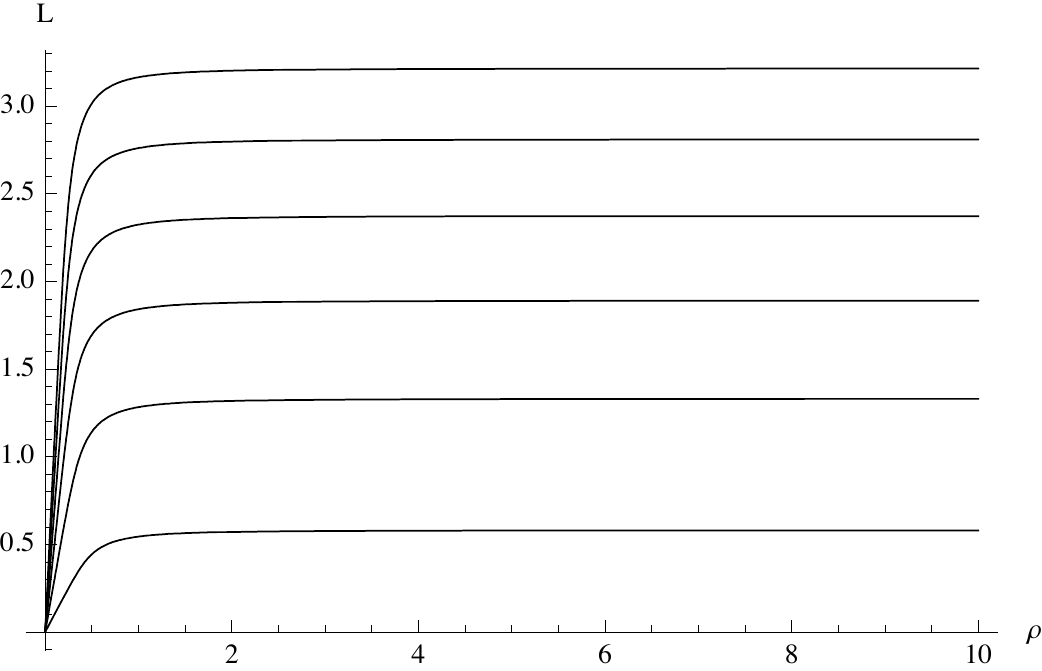}}
\caption{D7-brane embeddings in the case of pure $AdS_5\times S^5$ with finite baryon density. In this case the only dimensionless scale in the problem is the ratio of the baryon density to the quark mass.\label{fig: bardensads}
}
\end{center}
\end{figure}

For the case of pure $AdS_5\times S^5$ we illustrate the embeddings studied in \cite{Karch:2007br}. In this case the action for a D7-brane with finite baryon density is given by:
\be
S_{D7}=-\tau_7 \int dt d\rho \rho^3\sqrt{(1+\dot{L}^2+L^2\dot{\varphi}^2) -A_0'(\rho)^2}\, .
\ee
Again we use the $U(1)$ symmetry to set $\varphi=0$. We also note that there are two conserved charges as both $L$ and $A_0$ appear only with derivatives. The gauge field $A_0$ corresponds, as described above, to a finite baryon density. The conserved charge with respect to the gauge field is given by:
\begin{equation}
Q=-\frac{A_0'(\rho) \rho^3}{\sqrt{1-A_0'(\rho)^2+\dot{L}(r)^2}} \, .
\end{equation}
Performing the Legendre transform and replacing the gauge field with the conserved charge we are left with:
\be
\tilde{S}_{D7}=-\tau_7 \int dt d\rho\sqrt{(1+\dot{L}^2)(Q^2+\rho^6)} \, .
\ee
Note that there is a trivial scaling where we can rescale all dimensionful quantities either in units of $Q$ or in units of the asymptotic value of $L$ which gives the quark mass (or the non-normalisable behaviour giving the condensate). The physically relevant parameter we can tune is then $\frac{L_{UV}}{Q^\frac{1}{3}}$. Solving the equations of motion numerically and plotting the D7-brane embeddings in figure \ref{fig: bardensads} we see a family of solutions which have the clear throat-like behaviour in the IR. In the case of finite temperature there is an extra scale and so the possible phase space is more interesting, though the same general tendency holds that the fundamental strings dissolved on the D7-brane pull it into the horizon, giving the throat-like behaviour.


\subsection{Baryon vertex}

Having introduced finite baryon density in the previous section and realised that there may be a problem 
in the case of a confining geometry, we now turn to the study of the baryon vertex, first introduced by Witten 
in \cite{Witten:1998xy}.  The baryon vertex is a gauge invariant antisymmetric combination of $N$ external 
quarks, whose dual gravity description is given by a D5-brane wrapping a five-sphere at some radius in the $AdS$ space and connecting with the boundary by $N$ fundamental strings. (NB. we stick here to the case of AdS/CFT in the $AdS_5\times S^5$ context)

Witten's argument goes as follows:  In type IIB string theory there is a self-dual field strength $F_5$ 
and the compactification on $AdS_5 \times S^5$ provides $N_c$ units of flux on the five-sphere, 
$\int_{S^5} \frac{G_5}{2 \pi} = N_c$. On the D5-brane world volume there is a $U(1)$ gauge field 
$A$ which couples to the five-form field strength through the term $\int_{R\times S^5} A \wedge
\frac{G_5}{2 \pi}$. It is because of this coupling that $G_5$ contributes $N_c$ units of $U(1)$ charge. 
Each string endpoint adds $-1$ unit of charge and since in a compact space the total charge has to vanish, 
$N_c$ strings have to end on the D5 brane. In the $SU(N_c)$ gauge theory the gauge invariant combination of $N_c$ quarks is completely antisymmetric and, indeed, the strings between the boundary (or a D3-brane) and the D5-brane are fermionic strings.

There are two complementary approaches to the baryon vertex depending on the influence of the $N_c$ fundamental strings on the D5-brane. In the first one we neglect the deformation of the D5-brane and the world-volume gauge field on it, due to a uniform distribution of fundamental strings over the five sphere. In this way we treat the configuration as a combination of strings with a flat D-brane wrapping around the $S^5$, for the purpose of calculating the total energy of the brane \cite{Brandhuber:1998xy, Imamura:1998hf}. However, the distribution of $N_c$ fundamental strings breaks supersymmetry completely. Even if each string preserves one half of the supersymmetry, the fact that they take different positions on the D5-brane makes the preserved Killing spinor of each string different, such that all supersymmetry is broken \cite{Imamura:1998gk}.  

The supersymmetric solution is to have the D5-brane deformed by means of the tension of the strings attached to it. This deformation will happen if a significant number of them join at the same point. In this way the string back-reaction is not 
negligible and should be taken into account \cite{Imamura:1998hf, Imamura:1998gk}. In this case
the full DBI action should be considered, where the fundamental strings are seen as a spike in the 
world-volume of the D5-brane, along the lines of \cite{Callan:1998iq, Callan:1999zf}. 

In the approaches already discussed the quarks are non-dynamical and are defined by strings stretching from the D5-brane to the AdS boundary. In this way they are describing infinitely massive quarks. The question then is whether one can add dynamical (though not in the sense of unquenched) quarks with the usual probe embeddings of D7-branes and have the fundamental strings from the D5-brane attach to the D7's. On the gauge theory side (in the pure AdS limit) we will have ${\cal N}=2$ supersymmetry and fundamental quarks, so we expect dynamical, finite-energy, 
supersymmetric baryons. The gravity dual of such an object will be a D5-brane 
wrapping a five-sphere and connecting to the D7-brane with $N_c$ fundamental strings, 
\cite{Kruczenski:2003be}. The fact that the baryon vertex is supersymmetric (in the pure AdS background) is related to the orientation of the D7-brane. As long as it is parallel to the D3-branes and orthogonal to the strings the dynamical baryon will preserve half of the initial supersymmetry. The situation changes dramatically upon introducing a nonzero temperature. In this case, the gravitational attraction of the black hole pulls the D5-brane towards the event horizon until it collapses, leaving us with just a black hole embedding for the D7-branes, \cite{Mateos:2007vc}. This is to be expected as we already know that when we have a finite density of fundamental quarks dissolved in the brane it is pulled into the black hole under the tension of the strings, so the situation with and without the baryon vertex at finite temperature is identical. This is not surprising from the field theory point of view as in the deconfined phase we don't expect to have any baryons.

In the Sakai Sugimoto model the picture is very similar to the situation described above with a wrapped D4-brane mirroring the wrapped D5-brane in our description \cite{Hata:2007mb}. In this context the holographic dual to the skyrme model has been constructed whereby the baryon is built as a topological configuration of the pion degrees of freedom. The spectrum of baryons is then found by solving the Schrodinger equation in the potential which defines the moduli space of the skyrmion. An analogous picture has not been constructed in the current setup.


\subsection{Baryon density in a confining geometry}

At this stage we can motivate our setup before going onto the full construction in the following sections.

We know that if we add a finite baryon density to a D7-probe brane, this corresponds to having a density of fundamental strings dissolved in the brane world volume. The charges from these have to end somewhere and in the case of finite temperature the natural place for the strings to end is on the horizon. The baryon vertex can be thought of as a density of fundamental strings dissolved on the worldvolume of a wrapped D5-brane and these also need to end somewhere. In the trivial setup these end at the AdS boundary corresponding to a bunch of infinitely massive fundamental strings. However, if we want to introduce true quarks, corresponding to dynamical objects in the fundamental of a global symmetry and the fundamental of the color group these strings should end on a flavor brane. In a confining geometry it seems natural to connect the fundamental strings from the D5-brane with those from the D7-brane.

\begin{figure}[!ht]
\begin{center}
{\includegraphics[angle=0, width=0.6\textwidth]{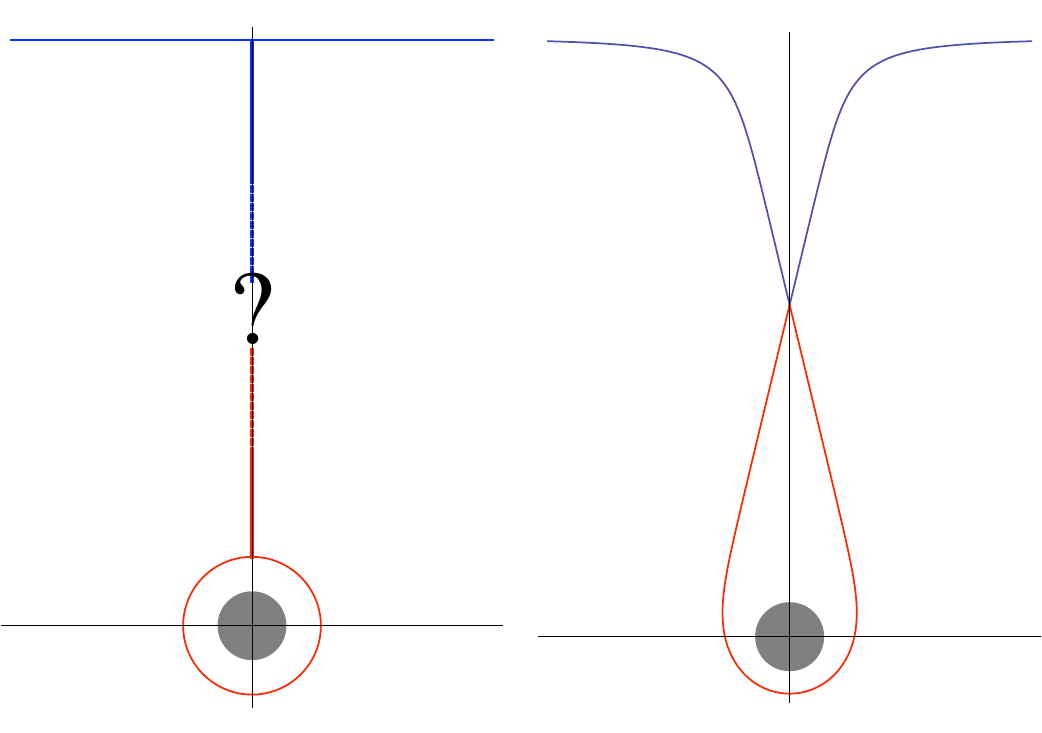}}
\caption{Left image: Blue D7-brane with a bunch of fundamental strings leading off it in the form of a throat plus red, wrapped D5-brane with a similar throat. Right image: The two throats meet at a point in the $(\rho,L)$ plane (the only allowed configuration in terms of the codimension of the objects). The configuration should be force balanced between the two objects. \label{fig:baryonvert}
}
\end{center}
\end{figure}

We can now motivate the computations which follow by a simple cartoon as shown in figure \ref{fig:baryonvert}.
The image on the left illustrates the most trivial, non-realistic setup. A D7-brane brane with finite charge density looks to have a throat made of fundamental strings which wants to end somewhere to carry off the charge. A baryon vertex made of a D5-brane wrapped on the five-sphere also has a throat made of a bunch of fundamental strings which wants to end on the AdS boundary. The natural interpolation (given that the fundamental strings pull the two objects together) is then the right diagram which shows the D5 and D7-brane meeting at a point, with an appropriate force balancing condition to create a stable situation. However, this may not be the true lowest energy solution but simply a metastable configuration.  It seems likely that given the point-like nature of the vertex there may be some non-trivial correction to this which will be the true vacuum. The case at hand is different from that of the Sakai-Sugimoto $D4-D8-\bar{D8}$ case as the D4-brane vertex is completely within the D8-brane worldvolume and therefore an instanton interpretation is natural. It would be interesting to know if there is an analogue to the skyrme model discussed in \cite{Hata:2007mb} in the present context. See section 6 in \cite{Kruczenski:2003be} for a discussion of the difficulties involved in such a calculation for the present case.


\section{Baryons in the dilaton flow geometry}


\subsection{D5-brane setup}

Here we study the embedding of a wrapped D5-brane in the GDFG. This will be the baryon vertex which, 
under force balancing conditions will attach at a point to the D7-brane, providing us with a baryon vertex 
where the constituents can be understood as dynamical quarks.

The background is given by the metric (\ref{ggmetric}) and we will set $r_0$ to 1 in all further numerical 
calculations. For the wrapped configuration of the D5-brane, the background five-form field strength (sourced by the D3-branes) will couple to the world volume gauge field $A_{0}$ via a Wess-Zumino term on the D5-brane worldvolume. The five-sphere coordinates are split into $\theta$ and $\Omega_4$ and we will find solutions where the scalar field on the D5-brane corresponding to the radial direction of AdS is a function of $\theta$ only. The gauge field will also be a function of this direction only. This leaves us with an $SO(5)$ symmetry for the wrapped, deformed D5-brane. We therefore have $r=r(\theta)$, $A_{0}=A_0(\theta)$. Such a setup corresponds to having the bunch of fundamental strings attached at a single point on the D5-brane world volume. The $r$ direction corresponds in the D7-brane language to $\sqrt{\rho^2+L^2}$. In the end the two solutions will join up at a point where $\rho=0$ meaning that $L=r$ at the vertex.

The induced, string frame, metric on the D5-brane with Euclidean signature is:
\be
ds_{D5}^2 = e^{\Phi/2}\left[ \frac{r^2}{R^2} A(r)^2 dt^2 +\frac{R^2}{r^2}\left(r'^2 +r^2\right)d\theta^2
+R^2 \sin^2\theta d\Omega_4^2 \right]\, ,
\ee
where $r'=dr/d\theta$. The DBI action for single D5-brane with $N_c$ fundamental strings can be written as:
\bea
S_{D5}&=& -\mu_5 \int d^6 \xi e^{-\Phi}\sqrt{{\rm det}(g+2\pi \a' F)}
+\mu_5 \int 2\pi \a' A \wedge G^{(5)} \cr
&=& \tau_5 \int dt d\theta \sin^4\theta \left[-\sqrt{e^{\Phi}A(r)^2(r'^2 +r^2)-\tilde{F}^2} +4\tilde{A}_0\right]\cr
&=& \tau_5 \int dt {\cal L}_{D5}\, ,
\eea
where we have denoted: 
\be
\tau_5 =\mu_5 \Omega_4 R^4 \, , \quad {\tilde A}_0=2\pi\a' A_0 \quad \& \quad \tilde{F}=\tilde{A}_0'(r)\, .
\ee
The dimensionless displacement can be defined as follows:
\bea
\frac{\partial {\cal L}_{D5}}{\partial \tilde{F}}
&=& \frac{ \sin^4\theta \tilde{F}}{\sqrt{e^{\Phi}A(r)^2(r'^2 +r^2)-\tilde{F}^2}}\cr
&\equiv& -D(\theta) \, ,
\eea
and the equation of motion for gauge field is given by:
\be
\partial_{\theta}D(\theta)=-4\sin^4 \theta\, .
\ee
Integrating, we have:
\be
D(\theta)=\frac{3}{2}(\nu\pi -\theta)+\frac{3}{2}\sin\theta\cos\theta +\sin^3\theta\cos\theta\, ,
\ee
where the integration constant $\nu$ determines the number of fundamental strings
($\nu \, N_c$ strings are attached to the south pole and $(1-\nu)\,N_c$ strings to
north pole). This solution holds true in any geometry where the $\mathbb{E}^6$ in the metric is not deformed (up to 
conformal rescalings). By performing a Legendre transformation with respect to the gauge field, we arrive at the Hamiltonian:
\bea\label{hamil-D5}
{\cal H}_{D5}/\tau_5 &=&\tilde{F}\frac{\partial {\cal L}_{D5}}{\partial \tilde{F}}-{\cal L}_{D5}\cr
&=& \int d\theta A(r)\sqrt{e^{\Phi}(r'^2 +r^2)}\sqrt{D(\theta)^2 +\sin^8\theta}\, ,
\eea
the equation of motion for which is:
\be\label{eom}
\frac{d}{d\theta}\left[\frac{r'A(r)e^{\Phi/2}\sqrt{D(\theta)^2 +\sin^8\theta}}{\sqrt{r'^2 +r^2}}\right]
-\frac{\partial}{\partial r}\left(A(r)e^{\Phi/2}\sqrt{r'^2 +r^2}\sqrt{D(\theta)^2 +\sin^8\theta}\right) =0 \, .
\ee
There is a trivial solution to this equation of motion given by $r=r_\star\sim1.471$ in units of $r_0$. This is the first artifact of the confining geometry. In pure AdS or AdS/Schwarschild there is no stable solution for a wrapped D5-brane as the minimum energy configuration is always when the brane collapses into the horizon (be it extremal or otherwise).

\begin{figure}[!ht]
\begin{center}
{\includegraphics[angle=0, width=0.4\textwidth]{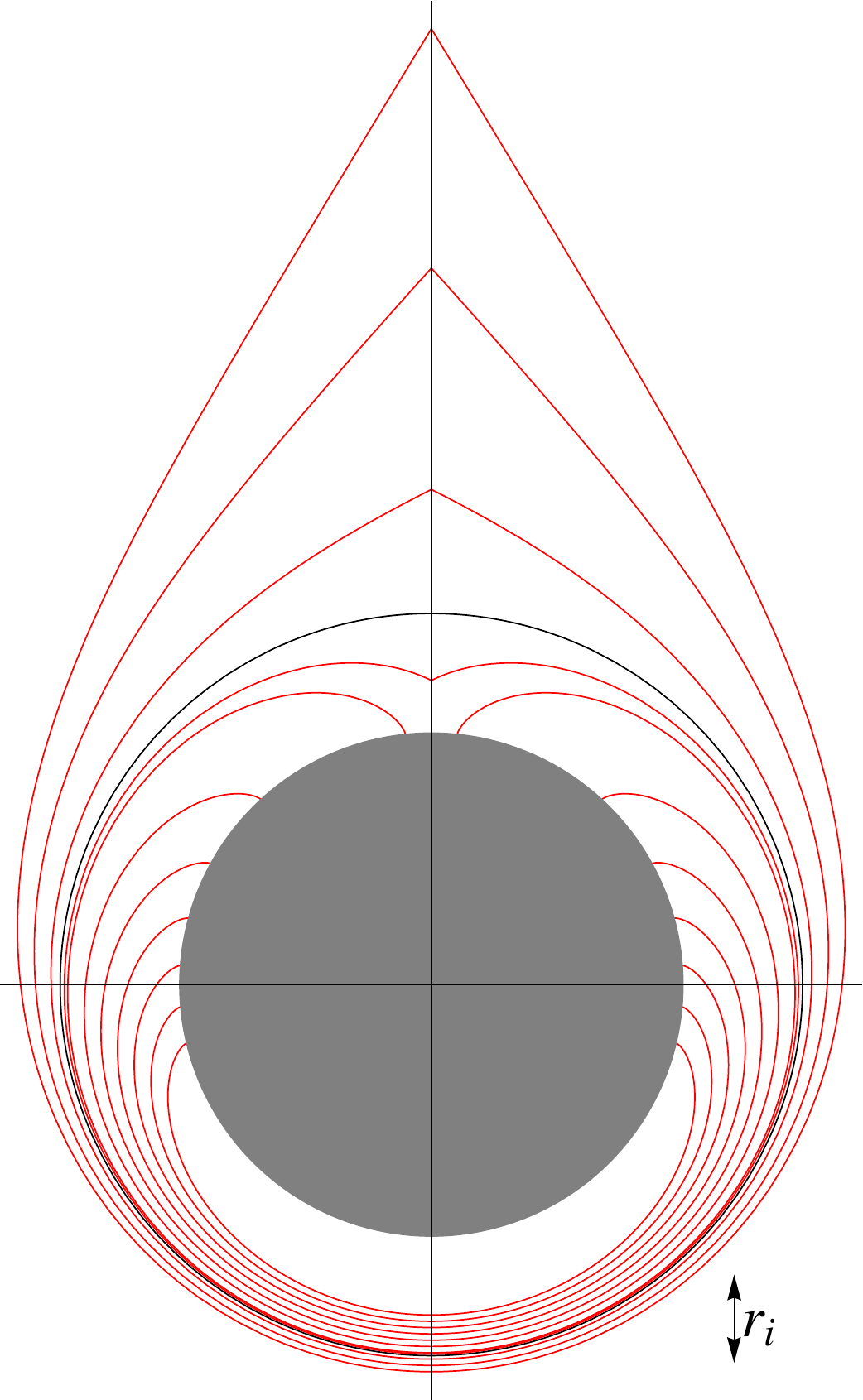}}
\caption{Wrapped D5-brane solutions for different $r_i$. The black circle corresponds to the $r_\star$ solution.\label{fig:baryon01}
}
\end{center}
\end{figure}

In the following we set $\nu=0$, meaning that all fundamental strings attach to one pole of the D5-brane, $\theta=\pi$. The equations of motion can be solved very simply using any numerical integration package. The boundary conditions which we impose are set at $\theta=0$. These are labelled $r_i=r(\theta=0)$, setting also $r'(\theta=0)=0$. Solving with these boundary conditions we find a range of different behaviours depending on the value of $r_i$. For small values of $r_i$ the D5-brane falls into the singularity, and clearly such solutions should not be trusted in the supergravity limit. For larger values of $r_i$ but less than $r_\star$ the solutions wrap around the singularity and end at $\theta=\pi$ at a value labelled $r_c$ with a positive gradient, $r'(\theta=\pi)<0$, whereas for solutions with $r_i>r_\star$ the solutions end at $r_c$ with $r'(\theta=\pi)>0$. This change in behaviour will be important in future sections. In figure \ref{fig:baryon01} we plot a range of different solutions including the singular solutions, and the 'well-behaved' solutions including, in black, the $r_\star$ solution.

\begin{figure}[!ht]
\begin{center}
{\includegraphics[angle=0, width=0.7\textwidth]{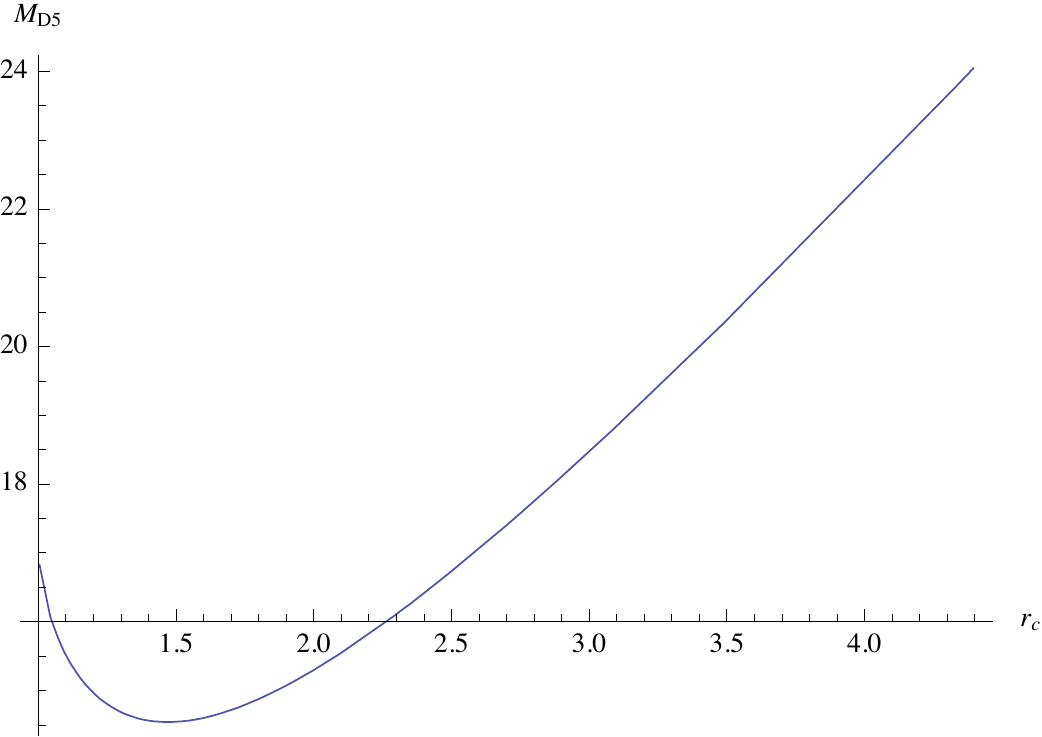}}
\caption{$r_c$ dependence of the energy of the wrapped D5-brane. The minimum gives the value $r_\star$ for which the pull of the singularity exactly cancels the pull of the fundamental degrees of freedom at the North pole of the brane.\label{fig:Md5_yc}
}
\end{center}
\end{figure}

The hamiltonian (\ref{hamil-D5}) gives the energy of the D5-brane. The $r_c$ dependence of the mass of a single D5-brane is drawn in figure {\ref{fig:Md5_yc}}. We see that as the value of $r_c$ goes to one, the mass increases but does not go to infinity as the brane touches the singularity (though the Hamiltonian density does diverge on the singular surface but slower than the worldvolume of the D5 vanishes). It should also be noted that in the absence of any external forces, the equilibrium position for the brane is at $r=r_\star$.\par

Because any D5-brane sitting away from $r=r_\star$ is out of equilibrium there is a force at the apex $r_c$ where the tension from the brane tries to pull it back to its equilibrium position. This force can be obtained from the variation of Hamiltonian of D5-brane with respect to $r_c$:
\bea
F_{D5} &=& -\frac{\partial {\cal H}_{D5}}{\partial r_c} \cr
&=& -N_c T_F \frac{A(r)r' e^{\Phi/2}}{\sqrt{r'^2 +r^2}}\Bigg|_{r=r_c}\, ,
\eea
where $T_F$ is the tension of a fundamental string and the direction is towards the singularity at the centre of the geometry (though note that for solutions with negative gradient $r_c<r_\star$ the force will be in the opposite direction). In order to match this force we can place another object which also has $N_c$ fundamental strings attached to it pulling in the opposite sense \cite{Seo:2008qc}. This is quite a natural set-up. The D5-brane has a bunch of fundamental strings attached in order to carry away the flux flowing through its world volume. We can let these strings go to the boundary in which case we can consider the system a baryon with infinitely massive quarks, or we can let them end on a flavor brane. The flavor brane which in this case is a D7-brane can have fundamental strings ending on it if we turn on a finite baryon density, as we have discussed above. Eventually the stable equilibrium of the system will be one where the fundamental strings have shrunk to zero size, pulling both the D5 and the D7-brane to a cusp.


\subsection{Probe D7-brane}

Now, we reconsider the D7-brane probe brane with finite baryon density in the confining geometry with the appropriate force balancing condition. The D7-brane solution will have a cusp where the fundamental strings from the D5-brane attach.
These end points can be understood as point charges on the D7-brane. 
As before, the action for $N_f$ D7-branes is:
\bea
S_{D7}&=& -N_f \mu_7 \int d\xi^8 e^{-\Phi}\sqrt{{\rm det}(g+2\pi \a' F)}\cr
&=& -\tau_7 \int dt d\rho A(r)^3 \rho^3 e^{\Phi/2}\sqrt{e^{\Phi} A(r)^2 (1+\dot{L}^2) -\tilde{F}^2}\cr
&=& \tau_7 \int dt {\cal L}_{D7} \, ,
\eea
where we have denoted:
\be
\tau_7 = \mu_7 \Omega_3 V_3 \quad \& \quad \tilde{F} = 2 \pi \a' F_{\rho 0} \, .
\ee
Just as in the pure AdS geometry we can look at the equation of motion for the gauge field to calculate the 
dimensionless quantity $\tilde{Q}$:
\be
\frac{\partial {\cal L}_{D7}}{\partial \tilde{F}}
=\frac{A(r)^3 \rho^3 e^{\Phi/2}\tilde{F}}{\sqrt{e^{\Phi}A(r)^2 (1+\dot{L}^2)-\tilde{F}^2}}
\equiv -\,Q \, T_F \equiv - \, \tilde{Q}\, ,
\ee
which can be replaced in the Legendre transformed hamiltonian to give us:
\bea\label{hamil-D7}
{\cal H}_{D7} &=& \tilde{F}\frac{\partial {\cal L}_{D7}}{\partial {\tilde F}} -  {\cal L}_{D7} \cr
&=& \int d\rho \sqrt{\tilde{Q}^2 + A(r)^6 \rho^6 e^{\Phi}}\sqrt{e^{\Phi}A(r)^2 (1+\dot{L}^2)} \, .
\eea
Just as before we can solve the equations of motion numerically, however now we have an extra force on the D7-brane coming from the fundamental strings stretching between it and the D5-brane. 
The force at the cusp of the D7-brane is equal to: 
\bea
F_{D7} &=& -\frac{\partial {\cal H}_{D7}}{\partial L_c}\cr
&=& - Q T_F \frac{A(r)\dot{L} e^{\Phi/2}}{\sqrt{1 +\dot{L}^2}}\Bigg|_{L=L_c} \, ,
\eea
where $L_c$ is the location of the cusp. This must be balanced with the force from the D5 and so we get the condition that:
\be
F_{D7} =\frac{Q}{N_c}F_{D5}\, ,
\ee
This states that the force on the D7-brane increases as we increase the quark density ($n_q\sim\frac{Q}
{N_c}$). The above relation can be written in the following simple form:
\be\label{fbc}
\dot{L_c} =\frac{r_c'}{L_c} \, , 
\ee
where we have used the fact that $r_c=L_c$. Now, solving the equations of motion with this as a boundary condition in the IR we can calculate the D7-brane embeddings for a given value of the quark mass and baryon density. In practice the calculation works in a different order: we pick a value of $r_i$ and solve the D5-brane embedding to find $r_c(r_i)$. Using this value, along with the behaviour of the D5 near the cusp as the boundary conditions for the D7-brane with a given baryon density we integrate this out to the boundary and read off the value of the quark mass. We  implement an efficient sequence of binary search routines to find an embedding with a fixed value of the quark mass as we vary $Q$. Now we can get various D7-brane embedding solutions with different choices of parameters.

\begin{figure}[!ht] 
\begin{center} 
{\includegraphics[angle=0, width=0.7\textwidth]{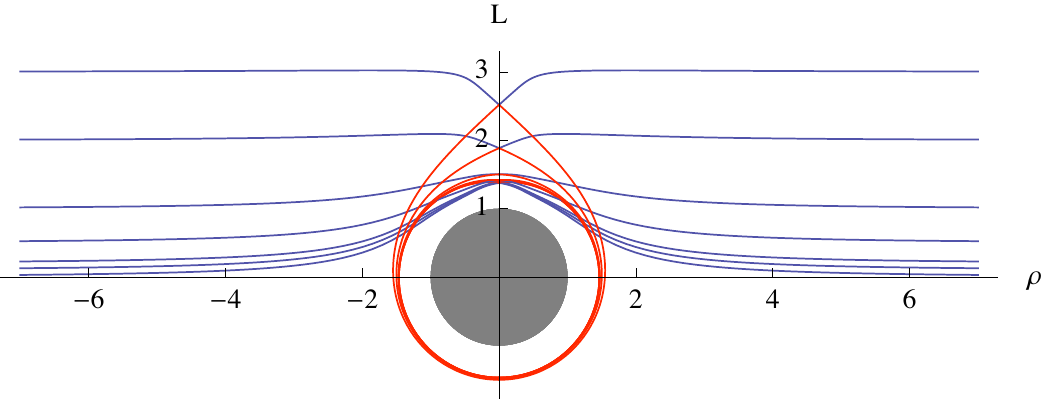}}
{\includegraphics[angle=0, width=0.7\textwidth]{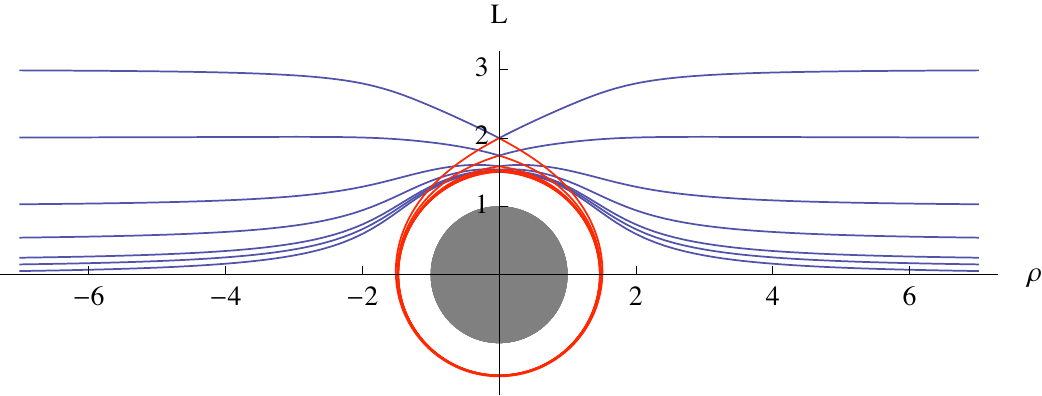}}
\caption{$m_q=0,...,3$ with (a) D7 and D5-brane embeddings for $Q$=0.1 (b) D7 and D5-brane embeddings for $Q$=5.
\label{fig:D7s-02}}
\end{center}
\end{figure}

\begin{figure}[!ht]
\begin{center}
{\includegraphics[angle=0, width=0.7\textwidth]{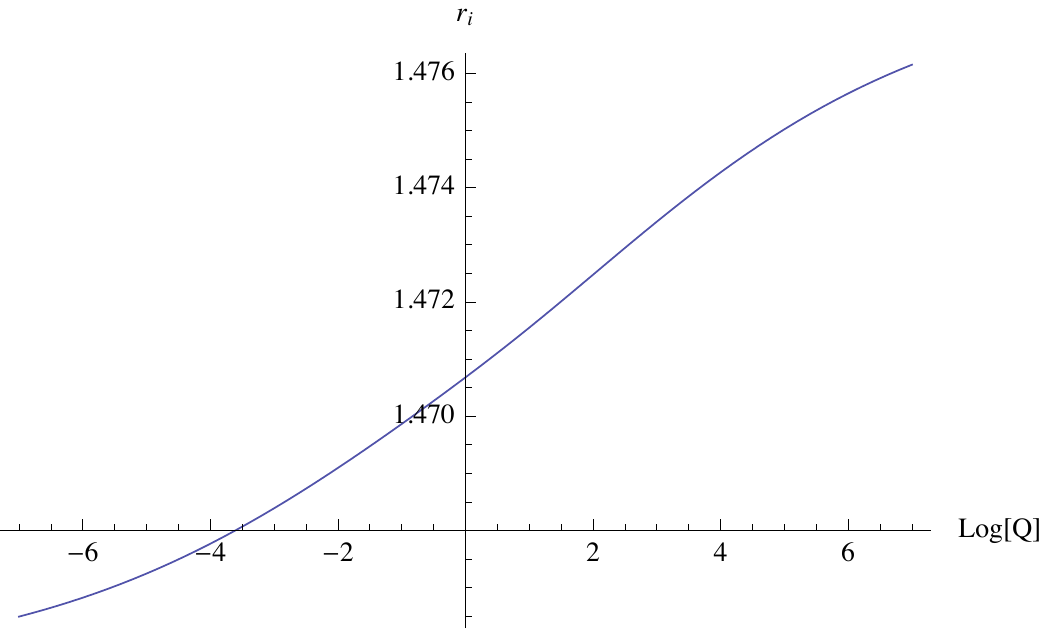}}
\caption{$Q$ dependence of the position $r_i$ of the South pole of the D5-brane. We see there is very slow dependence of $r_i$ on $Q$, however, as can be seen from figure 4, 
$r_c$ depends very strongly on the position of $r_i$. \label{fig:riQm0}
}
\end{center}
\end{figure}

In figure \ref{fig:D7s-02} we plot a range of $m_q$'s for different values of $Q$.
At each cusp at $\rho=0$ there is a balancing of forces between the D5-brane and D7-brane. It can be seen by comparing the two graphs in figure \ref{fig:D7s-02} that larger $Q$ affects the behaviour of the D7-brane from a larger value of $\rho$ whereas for small $Q$ the effect is only seen at small $\rho$.
It is interesting to see how we must change $r_i$, the South pole position of the D5-brane as a function of $Q$ for a fixed mass. In figure \ref{fig:riQm0} we plot this for the massless embeddings and note that there is a nearly logarithmic dependence on $Q$. Such a dependence will be seen in many plots which follow.

\begin{figure}[!ht]
\begin{center}
{\includegraphics[angle=0, width=0.75\textwidth]{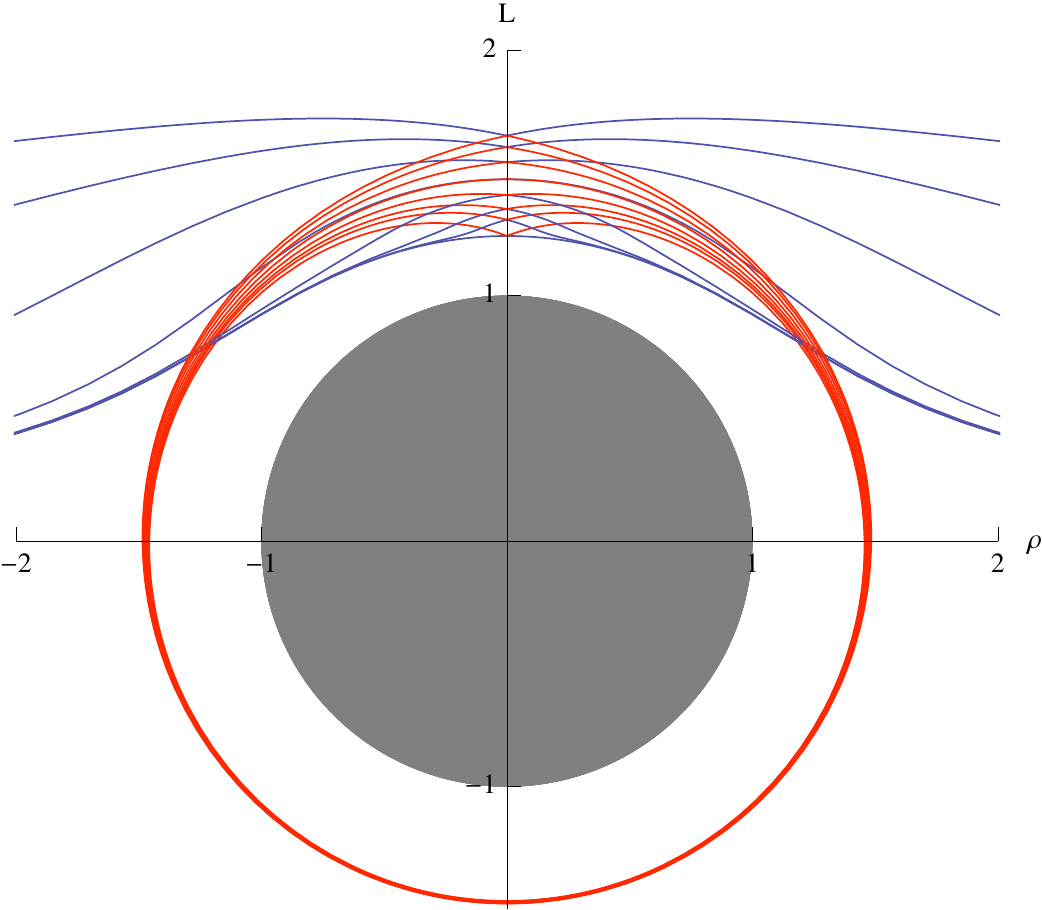}}
\caption{$m_q=0$ plots for a range of $Log[Q]$, from lowest D7-brane to highest $Log[Q]=(-\infty,-5.4,-3.4,-1.4,0.6,2.6,4.6,6.8)$ - D7-branes are blue, D5-branes are red.  We see that only for large values of $Q$ in this case $\gtrsim 10$ do the D5 and D7-branes not intersect each other away from the apex.\label{fig:m0largeQrange}
}
\end{center}
\end{figure}

\begin{figure}[!ht]
\begin{center}
{\includegraphics[angle=0, width=0.7\textwidth]{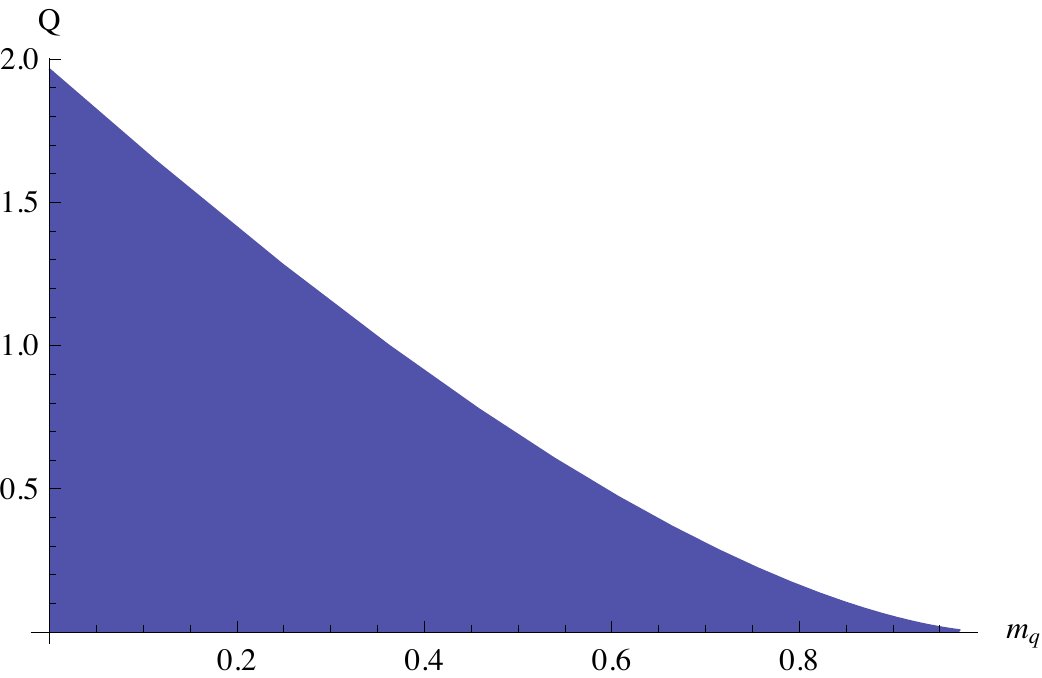}}
\caption{Region of intersections of the D5-brane and D7-brane shown in the blue shaded area below the line. The line corresponds to the embeddings with $r_i=r_\star$.\label{fig:posgrad}
}
\end{center}
\end{figure}

In figure  \ref{fig:m0largeQrange} we plot a series of $m_q=0$ embeddings with $Q$ ranging from $e^{-6}$  to $e^{6}$. We see that some of the D7-brane embeddings intersect with the D5-brane at a point other than the apex. Such solutions should perhaps not be trusted as there ought to be some non-trivial interaction between the two branes at these intersection points.
The question of whether or not the D5-brane intersects the D7-brane is a simple one to answer. Because of the force balancing condition the gradient of the D7-brane at the cusp is related up to a scaling, to the gradient of the D5-brane at the cusp. Only when the D5-brane has a positive curvature (in the coordinate system drawn in the figures here) will there be no intersection. The point at which this change occurs is at $r=r_\star$ and therefore we can parametrically plot the graph $(m_q(r_\star,Q),Q)$ (figure \ref{fig:posgrad}) which defines the line separating the region of intersecting solutions from the non-intersecting solutions.
It seems likely that the intersecting solutions are less physically realistic and so we will concentrate on the non-intersecting regions where possible. We do note however that the $Q\rightarrow 0$ limit recovers smoothly all known solutions in the absence of the baryon vertex and baryon density.

\begin{figure}[!ht]
\begin{center}
{\includegraphics[angle=0, width=0.9\textwidth]{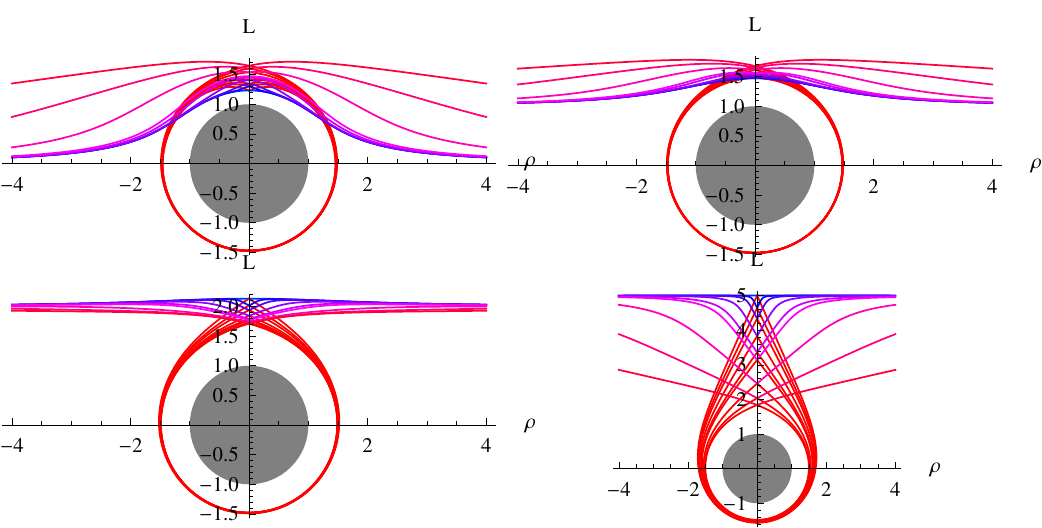}}
\caption{top left to bottom right: D5/D7-brane solutions for $m_q=(0,1,2,5)$ for $Q$ in the range $Log[Q]=(-\infty,-5.4,-3.4,-1.4,0.6,2.6,4.6,6.8)$. The D7-brane embeddings are color coded starting with blue for small $Q$ and going to pink for large $Q$. It can be seen that for smaller $m_q$, increasing baryon density lifts the value of $L_c$ whilst the opposite is true for larger $m_q$. \label{fig:plemb}
}
\end{center}
\end{figure}

In figure \ref{fig:plemb} we plot another set of D7-brane embeddings, but this time for fixed values of  $m_q$ in order to see the change in behaviour with $Q$. It is shown that for small and large quark masses, the effect of adding baryon density is different. For small $m_q$ turning on $Q$ increases the 'dynamical quark mass'  (increases the value $L$ at which the D7 and D5-branes meet), whilst for large quark mass the 'dynamical quark mass' is decreased. 
Note that the turnaround point where increasing $Q$ starts to drag the D7-brane down rather than up in the IR (for a given $m_q$) is around $m_q=1.6$.

\begin{figure}[!ht]
\begin{center}
{\includegraphics[angle=0, width=0.6\textwidth]{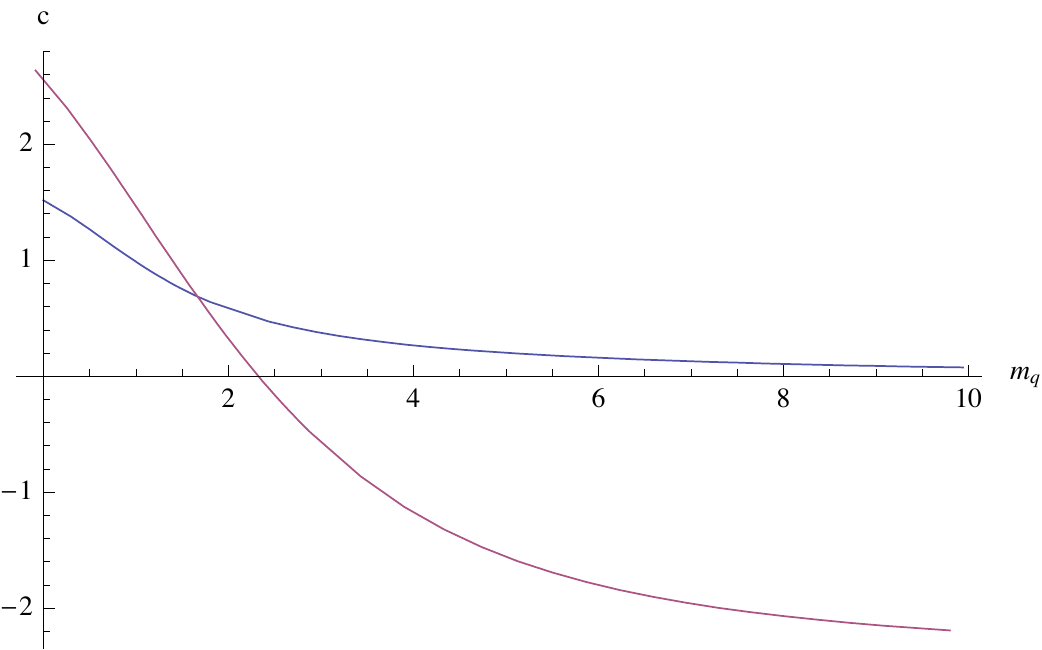}}
\caption{Chiral condensation for $Q=0.1$ (blue line) and for $Q=5$ (red line). Given that at finite baryon density in a non-confining geometry a chiral condensate with the opposite sign to that in a confining geometry at zero baryon density appears, a cross-over in behaviour is not surprising.\label{fig:chiral-01}
}
\end{center}
\end{figure}

For each of the sets of embeddings we can study the asymptotic behaviour of the D7-brane solution and, just as in the $Q=0$ case we can read off the value of the condensate. In figure \ref{fig:chiral-01} we plot the condensate as a function of the mass for two different values of $Q$.  We note that there is a chiral condensate for zero quark mass.



\begin{figure}[!ht]
\begin{center}
{\includegraphics[angle=0, width=1\textwidth]{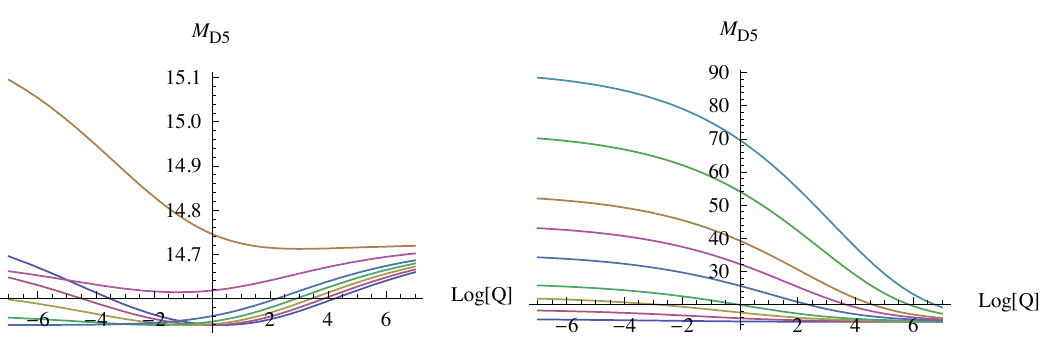}}
\caption{Density dependence of the D5-brane energy for different $m_q$. The left figure is (from bottom to top at $\log(Q)=7$) $m_q=(0.1,0.3,0.5,0.7,0.9,1.2,1.6)$ and on the right $m_q=(1.8,2.5,3.5,4.5,6,8,10,13,17)$. Again the very slow dependence on $Q$ is clear here.\label{fig:MD5logQ}
}
\end{center}
\end{figure}

Figure \ref{fig:MD5logQ} shows the $Q$ dependence of the energy of the D5-brane from the integral of the hamiltonian density for fixed $m_q$. 
We see in figure \ref{fig:MD5logQ} that for very small values of the quark mass there is a minimum value of the mass of the D5-brane which is absent in the D4/D6 system \cite{Seo:2008qc}. 


\section{Meson Spectrum}


\subsection{Setup}

At this point we turn our interest to studying the spectrum of mesons in our setup in the presence of a baryon vertex and baryon density. Having calculated the classical solutions we can ask about the excitations on top of the embeddings. The Minkowski-spacetime dependent excitations which have normalisable boundary behaviour in the UV correspond to excitations of the quark bilinear operators on top of any possible condensate value. These therefore correspond to mesons. Depending on the field we chose to excite on the D7-brane we will be able to study scalar and vector representations of mesons. The excitations are treated as small perturbations (of order $\alpha'\sim\frac{1}{\sqrt{\lambda}}$) and therefore we may expand the action up to quadratic order in these modes. Upon imposing the equations of motion for the background fields, the action linear in fluctuations vanishes and the equations of motion for the perturbations can be derived from the quadratic piece only. 

The fields which may be of interest in the subsequent analysis are the gauge field and the two 
scalars (corresponding to excitations of scalar and pseudoscalar quark bilinears). We make the gauge choice of $A_{\rho}=0$ and write our solution ansatz such that they depend only on $\rho$ and $x_0$. We label the fluctuations $A_{0,1,2,3}(\rho,x_0)$, $\tilde{L}(\rho,x_0)$ and $\tilde{\phi}(\rho,x_0)$. Clearly one could study the system where the fluctuations have finite momentum within the finite density medium, but this will unnecessarilly complicate the current analysis and so for the time being we concentrate only on finite frequency, zero momentum solutions.

The finite baryon density causes a coupling between the time component of the gauge field and the $\tilde{L}$ fluctuations. We are however most interested in the pseudoscalar $\tilde{\phi}$ which corresponds to the goldstone mode in the $Q=0$ case and this mode decouples completely even at finite baryon density. 
The action for this mode at quadratic order is given by:
\begin{equation}
{\cal L}_2(\tilde{\phi})\, =  {1 \over 2} \, \Big[ f(\rho)\,\partial_{\rho}\,\tilde{\phi} (\rho, x_0)^2
- h(\rho)\,\partial_{x_0}\,\tilde{\phi} (\rho, x_0)^2\Big] \, , 
\end{equation}
where we have denoted:
\begin{eqnarray}
f(\rho) &\equiv& e^{{1\over 2}\Phi(\rho)}A(\rho)  L(\rho)^2 \sqrt{\frac{Q^2+\rho^6 A(\rho)^6 
e^{\Phi(\rho)}} {1+ \dot{L}(\rho)^2}} \quad \quad  \& \nonumber \\  \nonumber \\
h(\rho) &\equiv& \frac{e^{{1\over 2}\Phi(\rho)} L(\rho)^2}{A(\rho)[\rho^2+L(\rho)^2]^2}\,\sqrt{(Q^2+\rho^6 
A(\rho)^6 e^{\Phi(\rho)})(1+ \dot{L}(\rho)^2)}  \, .
\end{eqnarray}
Calculating the equations of motion we use the following parametrization for the scalar:
\begin{equation}
\tilde{\phi} (\rho, x_0)= \varphi (\rho) \, e^{-i\omega x_0} \quad \rm{with} \quad \omega^2=M^2 \, , 
\end{equation}
and we arrive at the equation for the $\varphi (\rho) $:
\begin{equation} \label{scalar-eom1}
\partial_{\rho} \left[f(\rho) \partial_{\rho} \varphi (\rho) \right] + M^2 h(\rho) \varphi (\rho) =0 \, .
\end{equation}
It will be convenient to calculate the solution of the equation of motion after we transform 
it to a Schrodinger form, where we have found that the analysis is numerically more stable than in the canonical form\footnote{We will in fact use two Schrodinger forms in what follows, one of which is simpler for performing the numerical calculations and one of which gives a clearer picture of the bounded potential. The latter is introduced in the appendix.}.
In order to convert (\ref{scalar-eom1}) to a Schrodinger form we decompose $\varphi(\rho)$ as:
\begin{equation}
\varphi(\rho)\,=\,{1 \over \sqrt{f(\rho)}}\psi(\rho) \, , 
\end{equation}
and arrive at the following differential equation for $\psi(\rho) $:
\begin{equation}
\psi''-V_{eff}\psi=0 \, ,  \label{scalar-eom2}
\end{equation}
where the effective potential is given by:
\begin{equation} \label{eff-pot}
V_{eff}=\frac{1}{2}\frac{f''}{f}-\frac{1}{4}\frac{f'^2}{f^2}-M^2\frac{h}{f} \, . 
\end{equation}
Expression (\ref{eff-pot}), contains terms which have second and third derivatives in the background embedding. Such derivative terms are numerically very unreliable, so we remove them using the equation of motion for the embedding to write it purely in terms of zeroth and first derivatives.

In order to study the meson spectrum we study the excitations of the normalisable modes. 
Doing so we perform an asymptotic expansion of the equations in the UV and find that the potential 
and the corresponding solution in this limit behave as:
\begin{equation}
V^{eff}_{\rho\rightarrow\infty}\sim\frac{3}{4}\frac{1}{\rho^2} \quad \quad  \& \quad \quad 
\psi_{\rho\rightarrow \infty}\sim A\rho^\frac{3}{2}+\frac{B}{\sqrt{\rho}} \, .
\end{equation}
We therefore take the normalisable solution and set the UV boundary condition at some $\rho_{UV}$ to be $\psi(\rho_{UV})=(\rho_{UV})^{-1/2}$. In the IR we are supposed also to make sure that our wavefunction is normalisable. Since the potential goes to a negative constant, depending on the position of the D7-brane and its derivative, the solution behaves as:
\begin{equation}
V^{eff}_{\rho\rightarrow 0}\sim C \quad \quad  \& \quad \quad 
\psi_{\rho\rightarrow 0}\sim c_1 \cos\left[\sqrt{C}\rho\right]+c_2 \sin\left[\sqrt{C}\rho\right] \, ,
\end{equation}
$c_1$ and $c_2$ being integration constants. Calculating the integral of the square of the wave-function over the volume form it can be shown that both 
solutions give convergent IR behaviour and so are normalisable, meaning that we will have separate 
wave-functions with both Neumann ($N$) and Dirichlet ($D$) boundary conditions. In the case of D7-branes in pure AdS, 
$N$ and $D$ boundary conditions coincide. The IR behavior in pure AdS  is given by a 
sum of positive and negative exponential factors, so the normalisable solution has both vanishing value and derivative. In the current non-AdS case with finite $Q$, because of the two distinct 
boundary behaviours we may have a splitting in the spectrum (different spectra for $N$ and $D$ boundary conditions). This will correspond to a splitting in parity modes for the $0^+$ and $0^-$ mesons. In the case of zero baryon density where the wavefunction for the meson mode is even about $\rho=0$ the mode is pseudoscalar because of its transformation properties in the $(\rho,\phi)$ plane (see \cite{Kruczenski:2003uq} for a detailed discussion).


\subsection{Zero $m_q$ spectral flows}
\begin{figure}[!ht]
\begin{center}
{\includegraphics[angle=0, width=1\textwidth]{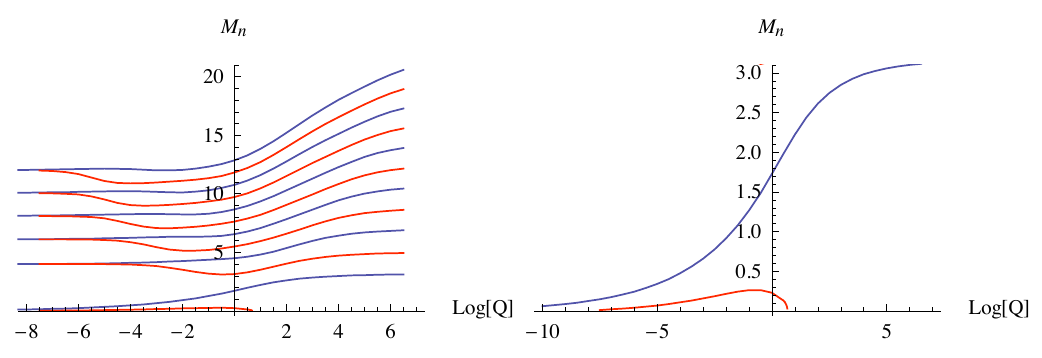}}
\caption{$m_q=0$ flows as a function of $Q$. We see clearly the splitting between the Neumann and Dirichlet modes and the tachyon in the lowest Neumann mode as the eigenvalue crosses zero (into the  complex $M_n$ plane). (Neumann in red, Dirichlet in blue). The Neumann flow in the right figure continues to smaller $Q$ (tending to $M_n\rightarrow 0$) but is not shown in this graph. The right plot is a zoom in on the small $Q$ region of the right plot.\label{m0flows}
}
\end{center}
\end{figure}

We now study how the baryon density in the confining geometry affects the goldstone spectrum. In order to do this we study the $m_q=0$ solutions as a function of $Q$. In the $Q=0$ case we have a massless pseudoscalar mode which follows a Gell-Mann-Oakes-Renner relation in the small $m_q$ limit. 
When we turn on a finite baryon density we find the spectrum shown in figure \ref{m0flows}.
There are several key features to this plot. The first bulk feature is that there is a splitting in the spectrum between $N$ and $D$ modes which coincide at $Q=0$. For large $Q$ the modes split and there seems to be an equal splitting between each mode.

There is also a clear difference in behaviour between $N$ and $D$ modes for small $Q$. The most important difference is that the $N$ modes contain a tachyon, indicating a clear instability above some value around $Q=2$. This tachyon, being in the $\bar{q}\gamma^5q$ part of the spectrum seems to indicate that the system wants to condense this operator. This would correspond to breaking of parity invariance (see \cite{Vafa:1984xg} for details about parity conservation in vector-like theories). However, it is believed that in QCD there might be such a parity breaking phase transition at high baryon density, it would be extremely interesting if we are seeing the signature of such a breaking in this context \cite{Andrianov:2007kz}. It seems that we are getting a splitting in parity pairs by the presence of the baryon vertex. Indeed the baryon vertex gives an explicit breaking of the chiral symmetry (therefore lifting the Goldstone mode) and allows for the splitting of the $0^+$ and $0^-$ modes.

\begin{figure}[!ht]
\begin{center}
{\includegraphics[angle=0, width=0.7\textwidth]{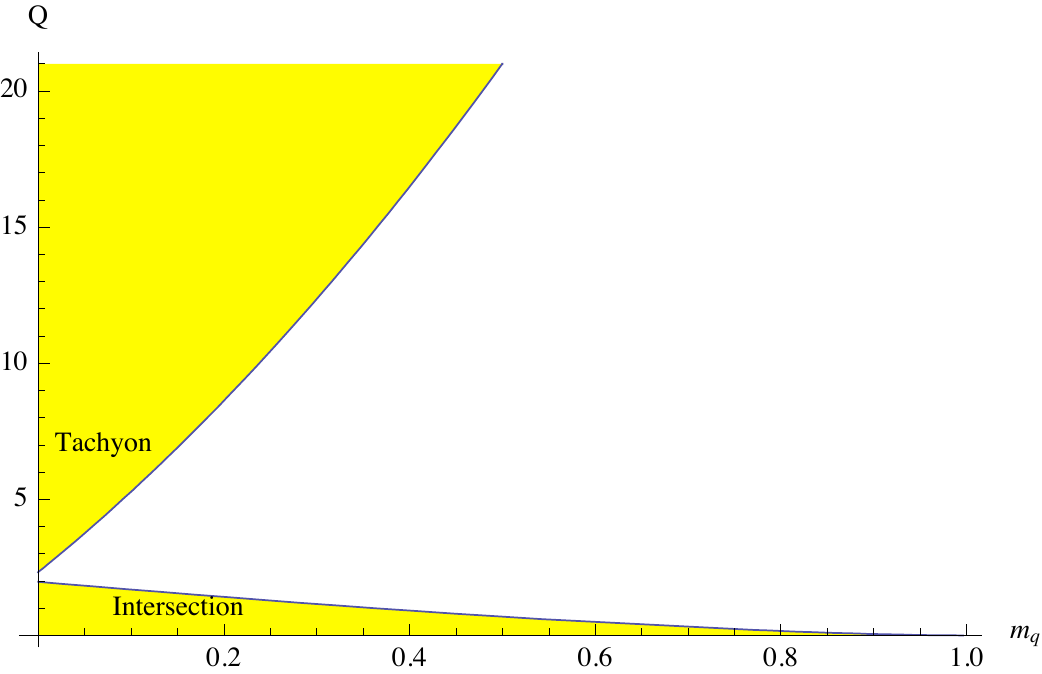}}
\caption{A plot showing the region with the tachyonic instability and the region with intersections of the D5 and D7-brane. The wedge in the middle is the only region which we believe to be stable from the current analysis. The regions with the intersection and the tachyon appear to coincide for $m_q=0$ at $Q\sim 2$.).\label{tachint}
}
\end{center}
\end{figure}

In fact the value of $Q$ for which we get the tachyon is about the same value of $Q$ where the intersections between the D5 and D7-brane disappear as the gradient of the D7-brane in the IR becomes positive definite. We can ask how the tachyon behaviour depends on the value of $m_q$ and this is plotted in figure \ref{tachint}.

\subsection{Large $m_q$ spectral flows}
We now turn to the limit of large $m_q$. In this limit at zero baryon density the D7-brane is unaffected by the deformation from pure $AdS$ and so the spectrum is known analytically to be $M_n=2 m_q\sqrt{(n+1)(n+2)}$. We can now follow the deformation from this result as we turn on the baryon density. We choose a quark mass $m_q=20$ which at $Q=0$ gives very good agreement with the pure $AdS$ result.

\begin{figure}[!ht]
\begin{center}
{\includegraphics[angle=0, width=0.7\textwidth]{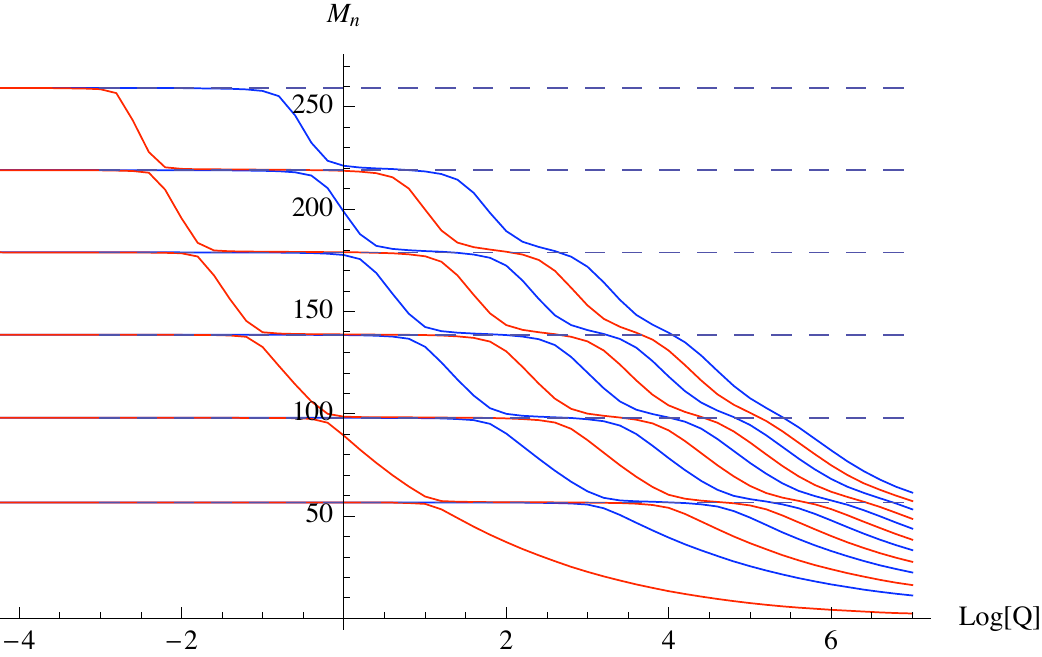}}
\caption{Spectral flow of the first six eigenstates as a function of $Log(Q)$. We see that the AdS solutions (given by dotted lines) act as attractors for the spectral flow, all the way to very large baryon density. As the value of a meson mass gets close to that of the pure AdS case, the wavefunction is only slightly deformed about the AdS dip in the potential. As the flow goes between these values the wavefunction is affected more by the second potential well and finally for very large values of $Q$ this is the dominant feature. Indeed we can continue this figure in the large $Q$ direction and find that the first mode becomes tachyonic at around $\log Q=9$ - again it should be noted that this very slow dependence on $Q$ is related to the dependence of the D5 embedding on $r_i$. Again, red is Neumann, Blue is Dirichlet.).\label{spectralflows}
}
\end{center}
\end{figure}

In figure \ref{spectralflows} we plot this spectrum and notice both some differences and similarities from the $m_q=0$ case.

The first similarity with the $m_q=0$ plot is that there is a breaking in the degeneracy between $N$ and $D$ modes. The clear difference is that the general trend of the spectrum is downwards in contrast to the $m_q=0$ case. However, this can be resolved by seeing the trend in the dynamical mass of the quarks as a function of $Q$ for large and small $m_q$ in figure \ref{fig:plemb}.	It should also be noted that there is a tachyon appearing at around $e^9$. This number seems very large, but because the units of $Q$ are $energy^3$ and the scale of $m_q$ is 20 in the current setup, the value of $e^9\sim 8000\sim 20^3$ is not unreasonable (Note that for small $m_q$ the other important scale in the problem is $r_0$ and so figure \ref{tachint} doesn't exhibit a cubic scaling for small $m_q$. Fitting the large $Q$ behaviour we find that roughly $M_n\sim \frac{m_q r_0}{Q^\frac{1}{3}}$.)

The other clear feature of the graph is the spectral flow whereby the spectrum moves between the pure AdS behaviour in steps. We can see this by looking at the Schrodinger potential in the form explained in the appendix in section \ref{sec.app1}. Note that this flow is extremely reminiscent of the spectral flow on the Higgs branch studied in \cite{Erdmenger:2005bj,Apreda:2006ie}.

\begin{figure}[!ht]
\begin{center}
{\includegraphics[angle=0, width=0.7\textwidth]{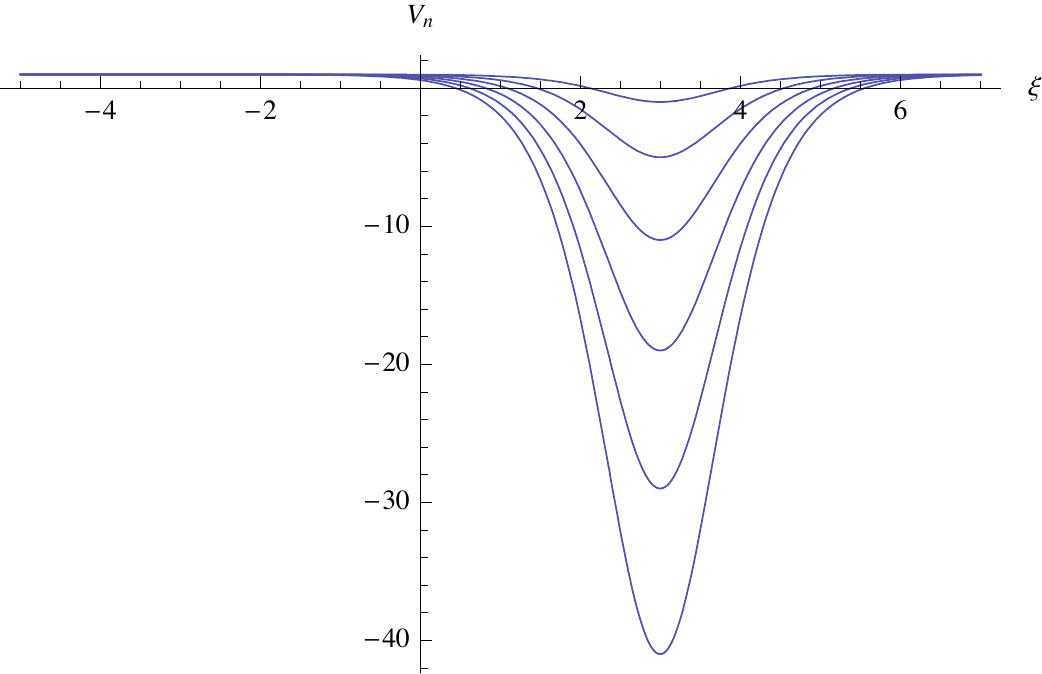}}
\caption{AdS potential for the first six meson masses for $m_q=20$.\label{AdSpot}
}
\end{center}
\end{figure}

In order to understand the behaviour we first study the Schrodinger potential for the case of pure AdS with zero baryon density. In this case there is a single minimum and the potential is given by the expression:
\begin{equation}
V_n=1-\frac{4 e^{2 \xi }m_q^2 (n+1) (n+2)}{\left(m_q^2+e^{2 \xi }\right)^2}\, ,
\end{equation}
This is plotted in figure \ref{AdSpot}.

\begin{figure}[!ht]
\begin{center}
{\includegraphics[angle=0, width=0.7\textwidth]{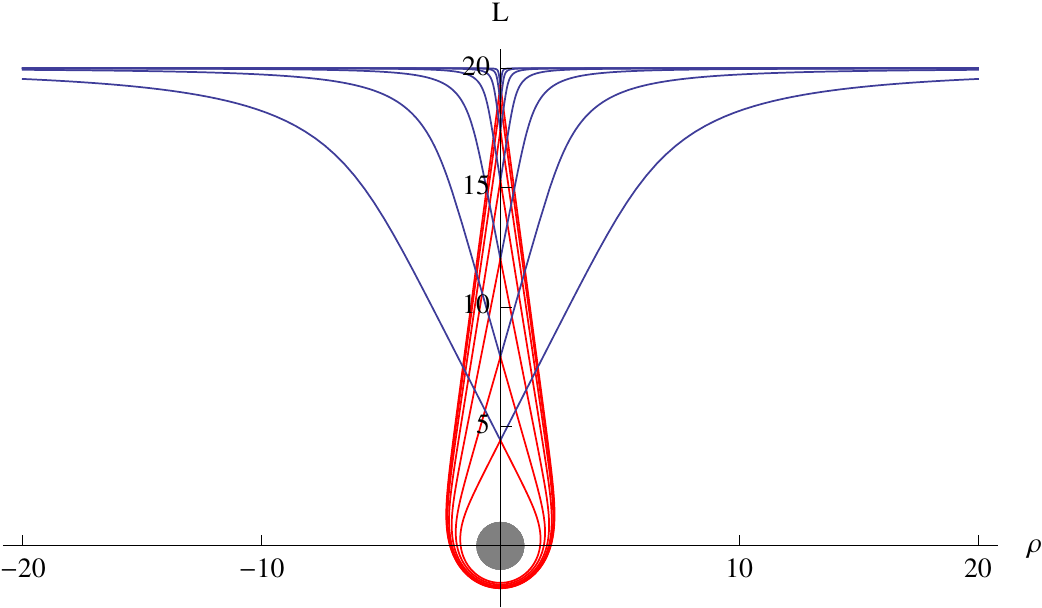}}
\caption{D5-D7-brane bound solution for $m_q=20$ and $\log(Q)=(-6,-4,-2,0,2,4,6)$.\label{D5D7mq20}
}
\end{center}
\end{figure}

On introducing a non-zero baryon density, the D7-brane is bound to connect with the baryon vertex and so we get a radical change in the behaviour of the potential. The change in the embeddings can be seen in figure \ref{D5D7mq20} illustrating the change in embeddings for a range of $Q$. We see that for small $Q$, the embeddings are less deformed from the $Q=0$ embedding and the deformation only appears at small $\rho$. The effects of this will only be observable for high energy solutions whose wavefunctions have non-trivial support mostly in this region. Such high energy solutions are the higher resonances of $M_n$. We therefore expect that as we increase $Q$ from zero, the higher resonances will be affected first, just as we see in figures \ref{spectralflows}.

\begin{figure}[!ht]
\begin{center}
{\includegraphics[angle=0, width=0.7\textwidth]{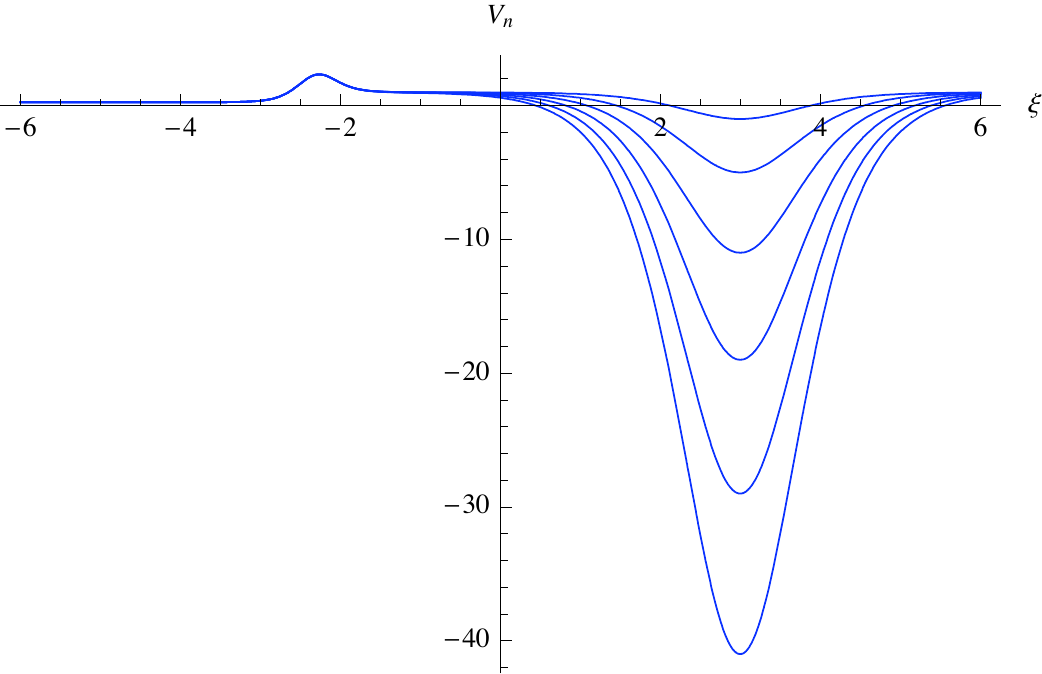}}
\caption{Schrodinger potential for the first six meson masses for an infinitesimally small baryon density. We see that the new feature is a very small perturbation on the AdS potential at around $\xi=-2$.\label{GubsersmallQ}
}
\end{center}
\end{figure}

The potential in the small $Q$ limit goes like:
\begin{equation}
V_{Q\rightarrow 0}=\frac{Q^4+32 e^{6 \xi } Q^2+4 e^{12 \xi }}{4 \left(Q^2+e^{6 \xi
   }\right)^2}-\frac{4 e^{2 \xi } m_q^2 (n+1)
   (n+2)}{\left(m_q^2+e^{2 \xi }\right)^2}\, .
\end{equation}
The IR limit of this has now changed from $V_{IR}=1$ to $V_{IR}=1/4$ and in accordance we have to alter the IR boundary conditions for any excitations. However, the dominant region is still the pure AdS potential well and so the spectrum remains the same as the AdS potential in the very small $Q$ limit (except for high energy resonances). The new feature of a maximum in the deep IR only affects these high energy solutions. We plot the potential for very small baryon density in figure \ref{GubsersmallQ}.
As we increase Q we see that the new maximum moves further into the UV, as the D7-brane is pulled more strongly towards the baryon vertex. We can see this from the brane embeddings, as shown in figure \ref{D5D7mq20}.

\begin{figure}[!ht]
\begin{center}
{\includegraphics[angle=0, width=0.7\textwidth]{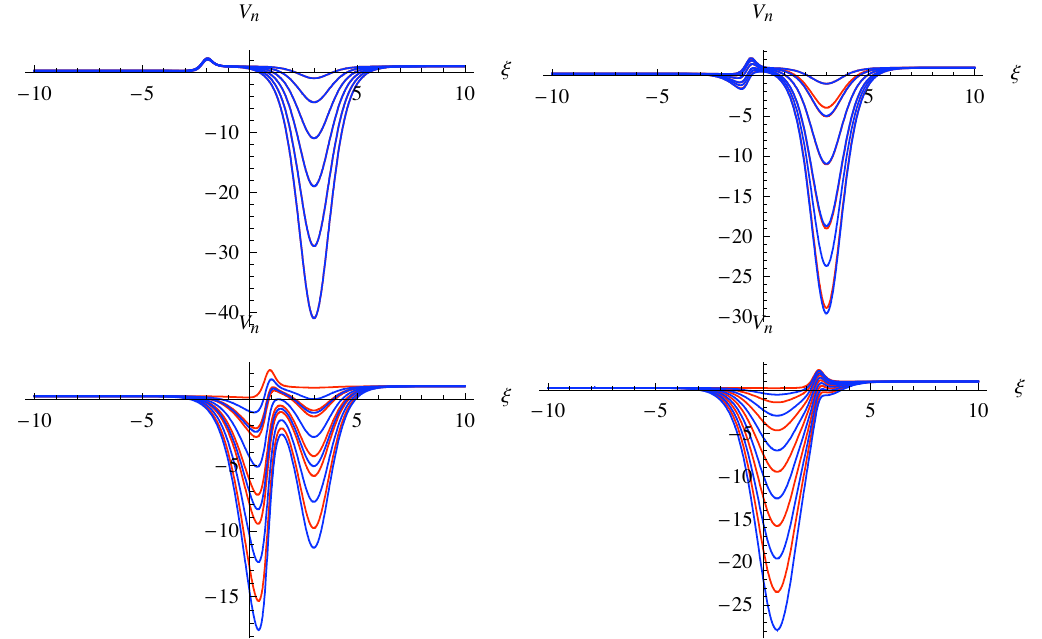}}
\caption{variation of the potential as a function of $Q$. Each plot contains the potentials of the first six eigenvalues and the four plots run from top left to bottom right with $\log(Q)=(-4,0,4,8)$. We see how the two competing potential wells dominate in different ranges of $Q$ values. (Neumann in red, Dirichlet in blue).\label{Qvariationofpotential}
}
\end{center}
\end{figure}

The evolution of the potential with Q involves two factors. The first is that the new feature in the potential moves further into the UV, and the second is that a new minimum appears next to the maximum which slowly becomes an important feature in the potential. In figure \ref{Qvariationofpotential} we plot the potentials for the first six modes for varying Q. 
If we plot the spectra along this flow in $Q$ as we traverse the behaviour between the single potential and the double potential form we see that the meson masses are most concentrated around the AdS behaviour, stepping quickly between the AdS eigenstates. As we increase Q, although the meson masses are going down, there is a step like behaviour and the $n^{th}$ eigenstate takes the value of the $n-m^{th}$ AdS eigenstate in $m$ steps until  we lose completely the AdS-like behaviour for very high values of $Q$. The reason for this step like behaviour is because as the dynamical quark mass decreases (the D7-brane is pulled further into the IR) with increasing baryon density, the spectrum decreases. Every time it gets close to the AdS spectrum it sits comfortably within the original AdS minimum and is not highly effected by the extra feature (the new minimum).

\section{Discussion}

In this paper, we have studied the meson spectrum in the presence of a finite 
baryon density in a confining gauge theory. The gravity dual of this gauge theory is a geometry with a flowing dilaton and a naked singularity giving a scale for a mass gap. The interaction between the fundamental degrees of freedom and the baryon density is 
encoded in the interaction of the flavor brane and baryon vertex, modelled by a wrapped D5-brane. 
We have examined how the meson spectrum flows as we increase the baryon density and found that for 
sufficiently large current quark mass, the mass spectrum decreases as the baryon density goes up. This is seen on the flavor brane by a decrease in the dynamical quark mass. For near zero current quark mass, the spectral curve does not show monotonic behavior. One of the most interesting phenomena discovered in the current work is the presence of a critical baryon density where the meson mass vanishes and subsequently becomes unstable. The presence of such a density is robust regardless of the current quark mass, though its magnitude  depends on the value of $m_q$. It would be interesting to study this further in the context of chiral symmetry restoration. Such an instability seems to signal the condensation of a pseudoscalar biquark operator though the true groundstate in this regime appears to be a more complicated interaction between the D7-brane and the D5-brane than we have been able to model. A natural extension to the current work would be to look for the stable solution for the large baryon density regime.

One of the new features in the calculation of the meson spectrum is that the IR boundary conditions for the D7-brane fluctuations are split into even and odd parity modes in the radial $AdS$ direction. This is brought about by the breaking of the chiral symmetry explicitly by the baryon vertex. It is clearly difficult to phenomenologically match these results with those of real QCD as experimental results at finite baryon density but zero temperature are not available. Such a situation would occur in cold, dense objects such as neutron stars and so any spectral signatures from such objects would be extremely interesting. 

In addition to the phenomenology studied here one could think of using the current setup as a model for more, realistic QCD-like phenomena. Clearly in a realistic dual of QCD (beyond the large $N_c$ limit) we expect to see mesons with a finite width. One possible way to simulate such a width without looking at large $N_c$ corrections to the geometry, would be to allow some transmission of energy from the D7-brane into the D5-brane through the vertex. To set up such a configuration however one would need to carefully tune the transmission coefficients of each meson mode such that there was a stable groundstate (in the absence of a dynamical photon). 

Another interesting point to pursue further would be the addition of a magnetic field in the set up 
along the lines of  \cite{Filev:2009xp, Erdmenger:2007bn}. In this case 
one could investigate its effect on the chiral symmetry breaking and the behaviour of the spectral curves. We leave such investigations for future studies.  


$\mathbf{Acknowledgements}$

We would like to thank Johannes Grosse, Javier Mas, Alexander Morisse, Andy O'Bannon, Alfonso Ramallo, Carlos Salgado, Javier Tarr\'\i o, Tatsuya Tokunaga and Patta Yogendran,  for helpful discussions throughout this project. The  work of JS and DZ was supported by the ME and FEDER (grant FPA2008-01838), by the Spanish Consolider-Ingenio 2010 Programme CPAN (CSD2007-00042), by Xunta de Galicia (Conselleria de Educacion and grants PGIDIT06 PXIB206185Pz and INCITE09 206 121 PR). J.S. has been supported by the ME of Spain by the Juan de la Cierva program.
YS was supported by the NRF  grant funded by the Korea government through the Center for Quantum Spacetime with grant number 2005-0049409. The work of SJS was supported by KOSEF Grant R01-2007-000-10214-0 and  also in part by the WCU project of Korean Ministry of Education, Science and Technology (R33-2008-000-10087-0).

\appendix


\section{Alternative form of the Schrodinger potential}\label{sec.app1}

Due to the numerical sensitivity in solving the eigenvalue problem we work with two forms of the Schrodinger potential, one of which is most useful for calculating the mass spectrum and is described in the body of the text, while the second, shown here, is more useful for indicating qualitative features in the spectrum from the point of view of the potential. In the following parametrization we start again from a differential equation of the form (\ref{scalar-eom1}) and transform it to a
Schrodinger like equation along the lines of \cite{Russo:1998by}.
The transformed equation will have the following form
\begin{equation}
\partial_\zeta^2\psi - V(\zeta)\psi = 0 \, , \label{schrr}
\end{equation}
and the necessary steps that bring it into this form, starting from  (\ref{scalar-eom1}), are 
\be
e^\zeta = \rho  \quad \quad \& \quad \quad \varphi = e^{\zeta \over 2} f^{-{1 \over 2}} \psi\ ,
\label{trass}
\ee
with the potential given by
\begin{equation}
V(\zeta) = -  M^2 {h_0 \over  f_0} + {1 \over 2} {f_0^{\prime\prime} \over f_0} -
{1\over 4} {{f_0^\prime}^2\over f_0^2} \ ,
\label{trass1}
\end{equation}
where 
\begin{equation}
f_0\equiv e^{-\zeta} f \ , \quad h_0 \equiv e^\zeta h\,  .
\end{equation}


\end{document}